**The Physics of Causation**

Leroy Cronin[1,2] and Sara I. Walker[2,3]

[1]School of Chemistry, The University of Glasgow, University Avenue, Glasgow G12 8QQ, UK.

[2]Santa Fe Institute, Santa Fe, New Mexico, USA.

[3]Beyond Center for Fundamental Concepts in Science, Arizona State University, Tempe, Arizona, USA

Emails: Lee.Cronin@glasgow.ac.uk; sara.i.walker@asu.edu

**Abstract**

Assembly theory (AT) introduces causation as a material property and establishes a metrology for objects produced by evolution and selection. The physical scale of causation is quantified by the assembly index, defined as the minimum number of recursive steps necessary to make an object. Observing countable copies of high assembly index objects indicates a mechanism producing them is persistent, such that the object's environment constructs a memory that traps causation within a contingent chain. Copy number and assembly index together underlie a standardized metrology for detecting causation (assembly index) and contingency (copy number). These allow a precise definition of an *assembly threshold* that demarcates life (and its derivative agential, intelligent, and technological forms and artifacts) as structures with persistent copies in regimes of deep causal possibility. In introducing a fundamental concept of material causation to quantify and measure life, AT represents a departure from prior theories of causation, such as interventional ones, which have so far proven incompatible with fundamental physics. We discuss how AT's concept of causation provides the foundation for a theory of physics that allows precise and testable concept of "life", and in which novelty, contingency and the potential for open-endedness are fundamental, and determinism is emergent from selection along assembled lineages.

## Introduction

Beyond a certain complexity threshold, the spontaneous formation of objects becomes vanishingly improbable, if not impossible. The chance of spontaneous formation for even moderately complex objects is small enough that this has never been observed. While spontaneous generation was ruled out as a plausible explanation for microbes over a century ago[1], the fluctuation into existence of arbitrarily complex objects persists as something allowed by current physics and introduces theoretical pathologies (e.g., Boltzmann brains, or requiring fine-tuning)[2,3,4]. A more parsimonious explanation is that objects must be caused to exist, but this is incompatible with current physics, which includes no concept of 'cause'.

Prior to the 1800s, life's origin was thought to either be spontaneous or co-eternal with non-living matter[5]. A prominent example was the observation that broth, even after boiling, if left in the open, would sprout micro-organisms[6]. It was even believed entire complex organisms could spring into existence instantaneously[7]. Careful experimental control of environmental and other causal influences has repeatedly demonstrated generating microbes is not spontaneous; for example, experiments with sterilized broth insulated from contamination exhibit zero growth of *de novo* complex, biological forms[1]. Likewise, many structures that chemists might envisage are so complex they will *never* be observed to emerge spontaneously within any span of finite time and finite resource. This is because the causation mediating their formation does not itself exist.

The observable universe is finite in time and resource. The age of our universe is ~13.8 billion years, and it is estimated to contain $\sim10^{80}$ atoms. Taxol, a secondary metabolite with molecular formula $C_{47}H_{51}N_1O_{14}$, shown in **Figure 1**, is one among an estimated $\sim10^{100}$ possible molecules[8] with molecular weight < 1000 Amu (Taxol's molecular weight is 853.91 Amu). Most of the $\sim10^{100}$ possibilities *will never* exist. There are no instantiated mechanisms (catalysts) for making the majority, and in any case, there are not enough atoms in the observable universe to make them all, nor is there enough time. For Taxol, the chance of producing it spontaneously, by joining 47 carbon, 51 hydrogen, 1 nitrogen and 14 oxygen atoms in the precise 3D arrangement of Taxol, is far smaller than 1 in $10^{23}$ (less than one molecule per mole of sample) with both the desired formula and 3D conformation.[8] The exact spontaneous probability cannot be defined: the size of chemical space is too vast and not computable, and it cannot ever be fully explored to assign probabilities to all possible structures within it.

**Figure 1**: Taxol's chemical structure with formula $C_{47}H_{51}N_{1}O_{14}$. The molecular weight is 853.91 Amu, and with 11 chiral centres there are $2^{11}$ or 2048 possible stereoisomers, a drastic underestimate of the total number of possible molecular structures with this same molecular formula.

Using ~$10^{100}$ as a lower bound on the number of chemical possibilities indicates the expected spontaneous molar concentration is so low that chemists should never observe Taxol, if it formed solely via spontaneous generation and an ergodic sampling of chemical space. In bark extracts from the Pacific yew tree, Taxol is abundant with a concentration ~20 ug $g^{-1}$ of bark (*ca.* $10^{16}$ molecules $g^{-1}$). Copies of Taxol are reliably synthesized as a secondary metabolite in Pacific yew bark, itself the product of billions of years of evolution constructing a specific organism that can make Taxol. For Taxol, both the biosynthetic, and also semi- and fully synthetic route for its synthesis are known[9–11].

How could we know Taxol must be the product of an evolutionary lineage, if we did not know anything about *how* it was made? This is one and the same as Paley's famous question about how one might determine design in a watch[12], but here applied to molecules as more tractable physical objects of study.

Paley intended his watch argument to be in favour of a theological explanation for design in nature. An alternative view, and the one we adopt herein, is that the appearance of design can be explained naturalistically, without a designer, via scientific methods: in essence, that the universe designs itself. As Krakauer has pointed out, explaining design encompasses many of the goals of the emergence of complexity[13] science in the last century (see early work by Simon[14]), which is unified across its many applications as the study of designed objects[15]. Darwin's "struggle for existence" introduced natural selection as an explanation for some features of organic design[16], via the mechanism of survival of the fittest. But natural selection is a passive filter, culling already-created forms to leave only the most persistent (adapted).

In modern evolutionary biology, the word 'selection' has been adopted to refer primarily to Darwinian processes operating on genomes. However, the Darwinian mechanism of selection is but one example of how selection can operate, and it is one that requires an already sophisticated evolutionary architecture (cellular life and genomes). Herein, we use ***selection*** in the more general sense of the definition of the word: the act of reducing a set of options to only those that persist over time. In this more general sense, one could consider selection as the fundamental mechanism that leads to the universe having any persistent structure (e.g. why is there something rather than nothing). At minimum we must assume selection is a process that can happen outside of biology; indeed, a mechanism of selection is necessary to give rise to biology and therefore must have been acting before the emergence of biology and its genomic architecture.

Open questions in evolutionary biology include why evolution discovered so many new, often complex, forms so rapidly[17]. While existing theories can explain aspects of speciation and phenotypic divergence, these are constrained within existing cellular architectures, levels of selection, and body plans. Yet historical records indicate unprecedented lineages can emerge, for example, in the evolution of cellular architecture at the time of the last universal common ancestor, and in the replication chemistry of viruses. Explanations for major transitions[18] are still open, and we do not fully understand how evolved information can branch into radically different substrates[3], such as when biology produces new technological forms. The two most significant transitions in the evolution of Earth are the emergence of biological lineages from geochemistry[19], and the emergence of technological lineages from biology[20], where both biological and technological lineages also have many subbranches[21].

Assembly theory (AT) introduces a standardizable, physical scale for constructed complexity by elevating causal possibilities to the status of a physical space, the assembly space[22], which can be metrologically explored[23]. Assembly space makes it possible to quantify a threshold, below which objects form spontaneously, and above which specific selective circumstances, the kind known only to arise through extant evolutionary or intelligent lineages, are necessary for an object to exist and be measured. Thus, AT provides a theoretical framework that allows experimental tests of whether objects like Taxol are objectively ones that *cannot* form spontaneously. The theory can be tested as a general theory of causation beyond molecules. This includes application to objects of biological or technological origin, or from another (alien) lineage.

To introduce the core concepts, we herein demonstrate the existence of a well-defined upper limit to what is possible in the absence of evolution and selection, relying only on a few, reasonable assumptions. We start with a theoretical proof of the *living threshold* in assembly space and discuss experimental verification. We use this as a segue into explaining AT's implications as a theory of physics, where causation is a measurable property of objects.

**Assembly Threshold Demarcating Life**

Thought experiments on the realizability of chemical space suggest the existence of objects cannot be taken as given *a priori*: we do not observe all possible objects (and never will). With assembly theory (AT) we introduce as a principle of physics that causation is necessary to the existence of objects.

In AT, each distinguishable *object*, $o_i$, is defined as a physical entity that is (1) countable, (2) finite, and (3) can be disassembled (or assembled) by a finite sequence of recursive steps (see Sharma et al.[22]). We define an **assembly index**, $a_i$, specific to each distinguishable object type, $i$, as the minimum number of causal steps necessary for that object to exist, where steps are built recursively. Here causation emerges in the application of physical constraints that must be present in the object's environment; an example is how catalysis can direct the formation of a chemical bond to make a specific molecular object, where the bond itself is the material evidence of causation.

Observing replicate **copies**, $n_i$, of an object indicates a mechanism is persistent enough to produce it reliably, which must be run from an extant memory (e.g., replaying the tape of assembly like an autocatalytic network depending on some catalysts in the environment). Any object persisting in environments for timescales longer than its natural half-life (timescale to decay), implies the object is constructed, and therefore can be constructed again[24]. This requires a mechanism was selected that itself persists in the environment, which allows the object to be reproduced faithfully (sometimes referred to in the abstract as a constructor, see e.g. von Neumann[25] or Deutsch[26]). This point is nuanced. For example, a molecule's half-life is controlled by its environment. If the environment includes conditions causing the molecule to react with other things (e.g., oxidation, reduction on Earth), this can cause the molecule to decompose over time. All organic matter will eventually decay to $CO_2$ and $H_2O$ because these are the thermodynamically most stable products[27]. Yet, via cellular replication, some molecular structures like ribosomes[28] have existed on Earth for billions of years. Deeper memory yields contingency and allows the production of copies of more intricate objects, that is, those with larger assembly indices.

Together, copy number and assembly index allow quantifying a threshold in assembly space, below which objects might form spontaneously, and above which specific selective circumstances, the kind known only to exist in extant evolutionary or intelligent lineages, are necessary for an object to persist and be measured. We therefore call this the ***assembly threshold***. To formalize our central argument, we consider two classes of objects elucidated by the example of Taxol: some objects are ***spontaneous*** because they can form in abundance in finite time without an external entity, and other objects are ***selected*** because they require very specific circumstances in their environment to mediate their formation in abundance in finite time, as a selective mechanism must compete against a combinatorially explosive set of possibilities. Selected objects must form along a ***lineage*** or causal chain specific to their formation. A key conjecture of AT is that this division is not only a feature of the processes making these objects, but of the objects themselves.

Our formal argument uses a very simplified mathematical example. Although simple, it demonstrates key principles of AT as a constructivist theory of physics, where no sufficiently high assembly index object can exist in the absence of a selective mechanism for its causal construction.

Assume a path that takes $d$ causal steps to construct an object (where $d$ is in assembly time, defined as the number of causal joins in a forward constructive process, see[22]). Here, a **causal step**, or causal join, is defined minimally, as in the joining of two existing structures to make a new one. In a combinatorial, recursively constructible space, the number of possible objects formed in successive causal joining events grows super-exponentially, see **Figure 2**.

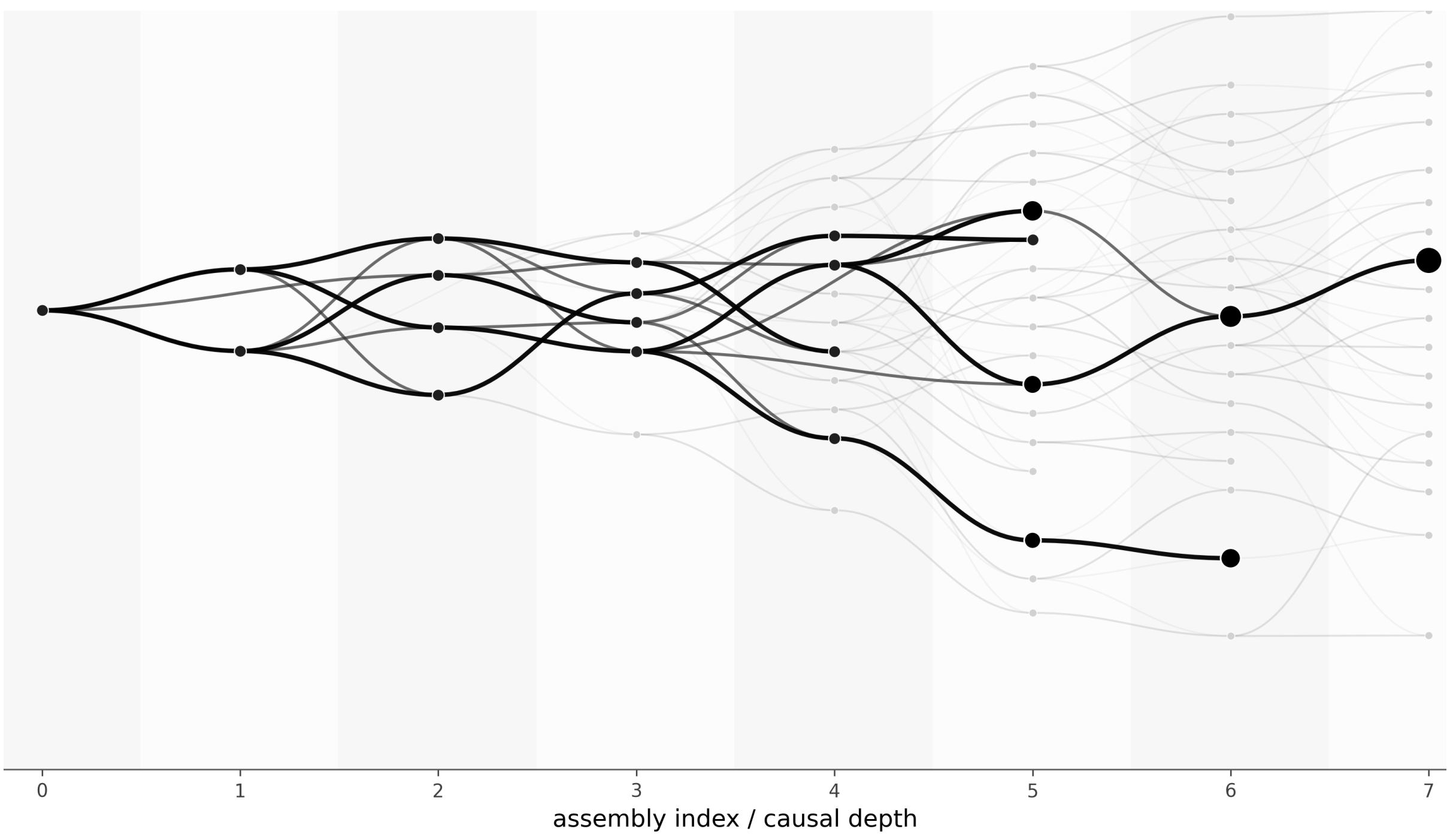

**Figure 2:** Branching of causal possibilities in assembly space. Light grey is assembly possible, black is assembly contingent, and black dots represent observed / persistent objects. Each possibility generated in a universe with causation will introduce a succession of new possibilities. Assuming two conditions (1) new possibilities are constructed from prior ones (recursively) and (2) new possibilities arise by minimum units of causation (pairwise joining of prior structures) yields the structure of the assembly space.

The constructable space of objects by recursive causal joins formally defines the assembly space (to be described later). We assume a branching factor, $b$, to parameterize the growth in the number of possible structures with each step. This branching will in general be substrate, and even object, specific. We define objects of type $i$ based on their indistinguishability, such that the number of copies, $n_i$, of a given type corresponds to the number of objects of that type which are indistinguishable up to limits of measurement. The **branching factor, $b_i$,** for a given object type, $i$, is the size of the set of other structures that can be formed in one recursive causal step, using only products of preceding causal joins along a causal chain. We simplify the mathematics with a global $b$, uniform across all objects for what follows to illustrate our key points. Due to this simplification, we expect our models to drastically underestimate the size of the assembly space. In physical samples we expect $b$ to increase with assembly index (e.g., taking a form $b_i(a_i)$): this leads to much more explosive combinatorial growth in the number of possibilities in real materials as physical systems traverse deeper regions of assembly space where object assembly indices are large.

Recursion is a necessary axiom for a physics that does not allow spontaneous generation of arbitrarily complex objects: this is because, in the absence of spontaneous existence, the only way to make new objects is to reuse preceding steps in new combinations along a causal chain. This also implicates microphysical causation as *necessary* to the physics we aim to describe. It follows that the exponential, combinatorial explosion of the space of possibilities with each step means there will always be a finite threshold for the causal depth of what objects can exist.

**Threshold Theorem:** A recursive causal threshold, defined herein as the *living threshold*, exists for all finite, combinatorial physical systems. Let $N_T$ be the total number of countable objects in a sample, $b$ the substrate-dependent branching factor (here taken globally for simplicity), and $M$ the instrument resolution defining the minimum number of copies of an object type necessary to register its measurement. We denote the threshold defining the maximum assembly index of an object that can be measured in multiple copies in the absence of selection as $a_M$. This is bounded by (see **Appendix A**):

**(1)**

$$a_M = \left\lfloor \frac{\ln\left(\frac{N_T}{M}\right)}{\ln(1+b)} \right\rfloor - 1$$

*Proof*: For a persistent object type to be measured, at least $M$ copies must be present in a sample. Assuming there exists a total number, $N_T$, of objects of any type, the expected number, $\langle n_i(d) \rangle$, of objects formed in $d$ steps is (**Appendix A**):

**(2)**

$$\langle n_i(d) \rangle = N_T \; e^{-(d+1)\ln(1+b)}$$

This form assumes all branches are followed uniformly, that is, there is no selection. The number density per object drops off exponentially with the number of steps needed create it. For finite $N_T$, there will be a branching point along every lineage where the expected number of objects drops to values $\langle n_i(a_M) \rangle < M$: the expansion of combinatorial possibilities is sufficiently fast that objects requiring more steps cannot persist above detection limits. Setting $\langle n(a_M) \rangle = M$, in Eq. 2, and solving for $a_M$ gives the threshold (see **Appendix A**).

This argument is sufficiently general that the behaviour of $a_M$ is not sensitive to the exact functional form of $n_i(d)$ and thus similar reasoning can apply to any recursively constructed combinatorial space constructed with finite resource: all such spaces will exhaust populating possibilities in a finite number of recursive steps, leading to a threshold in constructive depth, $a_M$, characteristic of the branching structure in the space. We do not weight by abiotic path probability because this introduces additional environmental assumptions beyond the central concern of whether an object is causally possible or not, and it does not change the existence of a living threshold (see Section on Possibility and Probability). In general, the parameterization of assembly steps, $d$, used in the forgoing calculations will not be strictly equivalent to assembly index, $a$. Our aim here is to build a model to exhaust all accessible possibilities such that *all objects emerging along all branches* persist and could be observed (representing exhaustive population of the possibilities). However, a forward assembly process can generate objects with symmetries that yield a minimum path shorter than the route taken (*s.t.* an object can have $a < d$). However, Eq. 1 nonetheless gives a fundamental bound on assembly index because the forward process can, by definition, never take a shorter causal route than the object's assembly index (by definition $d \nless a$). Eq. 1 thus demonstrates that, absent selection, *any* object we observe must be possible to form within a set of causal steps equal to $a_M$, or less, setting a fundamental causal bound on the possibility of an object to persist and be measured. Since assembly index measures the minimum causal steps to form an object, the value of $a_M$ tells us what objects *cannot* form without selection: object types with $a_i > a_M$ cannot form abiotically. Importantly, such a bound must hold if microphysical causation is a real feature of our universe; that is, if all objects have a partial ordering on what can exist (e.g., an object cannot be assembled unless the parts to assemble it are themselves already assembled, rendering spontaneous fluctuation impossible unless it follows a causal assembly path), see **Figure 3.**

By the foregoing, in a sample of $N_T$ abiotic objects, those with an assembly index $a_i > a_M$ will have an expected copy number $n_i \ll M$, meaning they cannot be measured or will not exist without some prior constraints in the branching process, e.g., being promoted by some form of selection along causal chains as could occur via chemical evolution. A key prediction of assembly theory is that any such object is a living artifact (either alive or the product of life).

The parameter $M$ places $a_M$ as an *epistemological limit*, set by the resolution of the measuring devices. In the limit $M \rightarrow 1$, where the resolution of measurement is *exactly one object*, the threshold $a_M$ is identical to an *ontological limit*, $a_O$. This captures the maximum expected assembly index in the absence of life for the *existence* of any object, defined where $\langle n(a_O) \rangle = 1$ in the absence of selection. For most systems of interest, due to the exponential size of the assembly space, the epistemological and ontological limits will be quite close (see e.g., **Figure 6**).

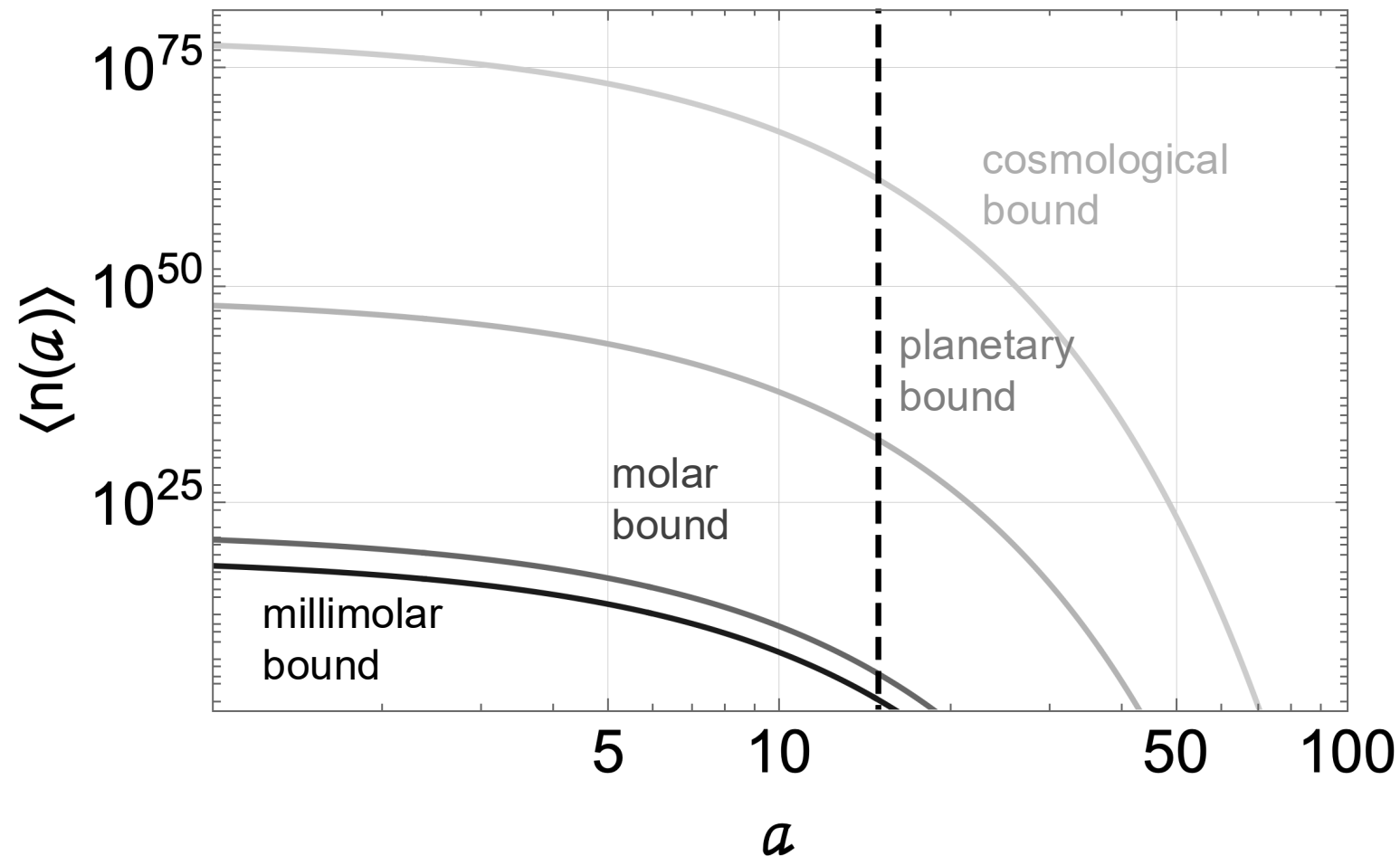

**Figure 3:** The expected copy number, $\langle n_i(a) \rangle$, of objects at a given number of steps, $a$, for varying volumes of total material, provides an upper bound on assembly index in the absence of selection.

The branching factor satisfies $b \geq 1$, where the limiting case $b \rightarrow 1$ occurs only if each object can undergo only one causal join and can make only one new possible object. In chemical space, $b$ is expected to scale with $a_i$, and the number of possible molecules is expected to scale super-exponentially with $a_i$; thus, the bounds presented herein are conservative as they significantly underestimate the combinatorics. Experimental data, as well as meteorite and astronomical data confirm only molecules with small $a_i$ are found abiotically[23], thus we expect the constant $b$ to provide a reasonable first approximation to the combinatorics of assembly space in the absence of life.

The ontological threshold, $a_O$, gives a natural division between spontaneous and selected objects, with important implications:

1. Objects with assembly index, $a_i$, below the threshold ($a_i < a_O$) can arise spontaneously in finite time
2. Objects with assembly index above the threshold ($a_i > a_O$) require specific causal histories and *cannot* form in abundance in finite time without being promoted by selection.
3. Observing high copy numbers of high assembly objects ($a_i > a_O$, $n_i > M$) is a signature of selection along a deep causal lineage.

The threshold is weakly system-size dependent, scaling logarithmically with sample size (the number of objects $N_T$), see **Figure 3** (e.g., it does not change by even one order of magnitude going from laboratory to cosmological scales). The dependence on $N_T$ renders $a_M$ an *extensive physical variable*, which is both substrate dependent (via branching, $b$) and volume dependent (via system size, $N_T$). This allows setting bounds on $a_M$ for laboratory samples, and using this to determine planetary and cosmological bounds, see **Figure 3**. For the curves shown in Figure 3, the total number of objects is $N_T = 10^{20}$(1 mmol), $N_T = 10^{23}$ (1 mole), $N_T = 10^{50}$ (approximate number of atoms composing the Earth), $N_T = 10^{80}$ (approximate number of atoms in the observable universe). The value of the branching factor, $b$, was estimated using the experimentally determined abiotic upper bound of $a_M \sim 13$ identified in Marshall et al.[29] (dashed line), where no molecule with an assembly index $a_i > 13$ was detected in millimolar samples of abiotically-derived material, with a detection limit of $M \sim 10{,}000$ for the mass spectrometer used. Using Eq 1. with $a_M = 13$, $N_T = 10^{20}$ (~1 mmol), and $M = 10{,}000$ yields an approximate averaged global branching over small molecule organic chemical space of $b \sim 12$ from our simple model. Setting $b = 12$, we can extrapolate to different volumes of material to determine the scaling of a threshold value of $a_M$ where the ontological bound is achieved, that is where $n(a_O) = 1$. For a molar sample of abiotic material $a_O(10^{23}) \sim 20$, the planetary bound is $a_O(10^{50}) \sim 44$ for an abiotic system with roughly as many molecules as there are atoms on Earth, and the cosmological bound is $a_O(10^{80}) \sim 71$ for an abiotic system with roughly as many molecules as there are atoms in the observable universe. Although these are very rough estimates, this demonstrates the existence of a cosmological bound: it is possible to extrapolate from laboratory measurements to make universal predictions, where *any* molecule with $a_i > a_O(10^{80}) \sim 71$ must be signatures of deep causal lineages; these cannot form spontaneously even leveraging all atoms in the observable universe. We expect this to be an overestimate, in particular because (1) the assembly space grows much faster than we account for here, (2) the number of molecules in the universe will be smaller than the number of atoms, with the number that can be made given a finite set of atoms decreasing with increasing molecular weight, (3) the mass in the model is distributed over the entire universe so the density for interaction is much lower than laboratory samples and (4) the experimentally determined millimolar bound of ~13 may be an overestimate based on limited analytical capabilities (e.g. peak counting[23,30]) such that the threshold could be lower. Future work can include more realistic assumptions, and more experimental data, leading to tighter bounds and predictions from the theory about the precise location of the threshold and how it scales in different materials.

Our example here is intended to be simple enough to show a critical role for both assembly index and copy number, combined with an ontology of microphysical causation, as necessary to defining a universal threshold for detection of the constructive causal lineages more colloquially referred to as 'life' and/or 'intelligence'.

**Assembly Space**

The above argument for a threshold demarcating living lineages is well-defined within assembly space, which encodes causal possibilities as a physical space that can be metrologically explored. In AT, objects exist with coordinates (assembly index, copy number) in the assembly space just as objects exist with coordinates in spacetime in more familiar physics; that is in assembly theory causation and possibility are physically real. Assembly spaces are substrate specific[23,31–33], which we shall discuss in more detail later, here we introduce a general definition, following that in Seet et al.[34].

**Definition 1 - Assembly Space**. An assembly space $(\Omega, U, J)$ consists of, $\Omega$, its set of possible objects, $U \subset \Omega$, the set of elementary units for assembling objects, and possible causal joins $J: \Omega \times \Omega \times \Omega \rightarrow \{0,1\}$, where $J(x, y, z) = 1$ indicates the causal join can happen such that objects $x$ and $y$ can be assembled to form object $z$, e.g., $\{x, y\} \rightarrow z$. The assembly space $\Omega$ is the closure of $U$ under all joining operations $J$, meaning $\Omega$ contains all objects that can be constructed from the set of basic units $U$ through any recursive sequence of possible causal joins. For combinatorial, compositional spaces, like chemical space, languages, memes, and morphologies, and for defined units and joining operations, the assembly space cannot be iterated using finite resource and finite time.

All information that can be known about the structure of assembly space is determined by deconstructing (fragmenting) observed objects. Thus, the assembly space of observed objects encodes information about a much larger set of potential possibilities, allowing large volumes of possibility to exist in small volumes of space and time, e.g. in the form of evolved objects[3]. Assembly spaces derived from observed objects will define all other possible objects constructible from the basic units and recursive causal joining steps[29], and all causal pathways connecting these possibilities. The assembly space constitutes what structure the universe can create that can be anticipated and is computable because the causal trace of these possibilities is present in existing objects. However, this is not everything that could be caused to exist. We aim to describe the physics of a creative universe where genuine, unanticipated novelty can happen, which is described in how the assembly space itself is also expanding and this cannot be computed, nor predicted in advance (see Section Open-Ended Evolution, Novelty and the Mechanism of Selection).

The assembly space has several important properties. It is combinatorially constructed, such that structures can combine in many ways. It is recursive, such that objects are only possible to assemble with the products of preceding steps along causal chains. These features establish assembly space as defining a *causal possibility space* where both causal processes and the objects they generate are encoded as the same structure. This provides a unified view of dynamics (laws) and objects, which have traditionally been considered distinct in other physical ontologies. See, for example, Walker and Davies[35] for a discussion of a need to unify states and dynamical laws to arrive at a physics descriptive of life. AT collapses the distinction between concepts of laws and states: the structure of the assembly space makes no distinction between sequences of causal joins and the objects to be measured: objects are recursive causal chains.

Assembly spaces can be described in nested layers[22], the exact shape and size of which will depend on how physical constraints are encoded in the joining operations $J$. The outermost layer, the Assembly Universe, $A_U$, is the least constrained and includes logical possibilities not exclusive to the physically possible ones, e.g., this set contains the mathematically imagined set of all possible assembled combinations. Defining physically implementable joining operations restricts the space to a set of causally possible objects, the Assembly Possible, $A_P$. However, even this set is far vaster than anything that could exist or be observed. Within the set $A_P$, are Assembly Contingent, $A_C$, spaces: these are spaces that can exist, constructed by selection in finite time with finite resource. Assembly Observed, $A_O$, is one such contingent space and describes the world we observe. While only $A_O$ can be directly interrogated and measured, its existence encodes the potential existence of other contingent spaces and the entire $A_P$ and $A_U$. While numerically it may appear that $A_U$ is the largest, and $A_O$ the smallest of these spaces, we regard the real physical "size" as inverted. What exists is $A_O$ : the possibility of $A_P$ and $A_U$ exist within observed objects in $A_O$, see **Figure 4**.

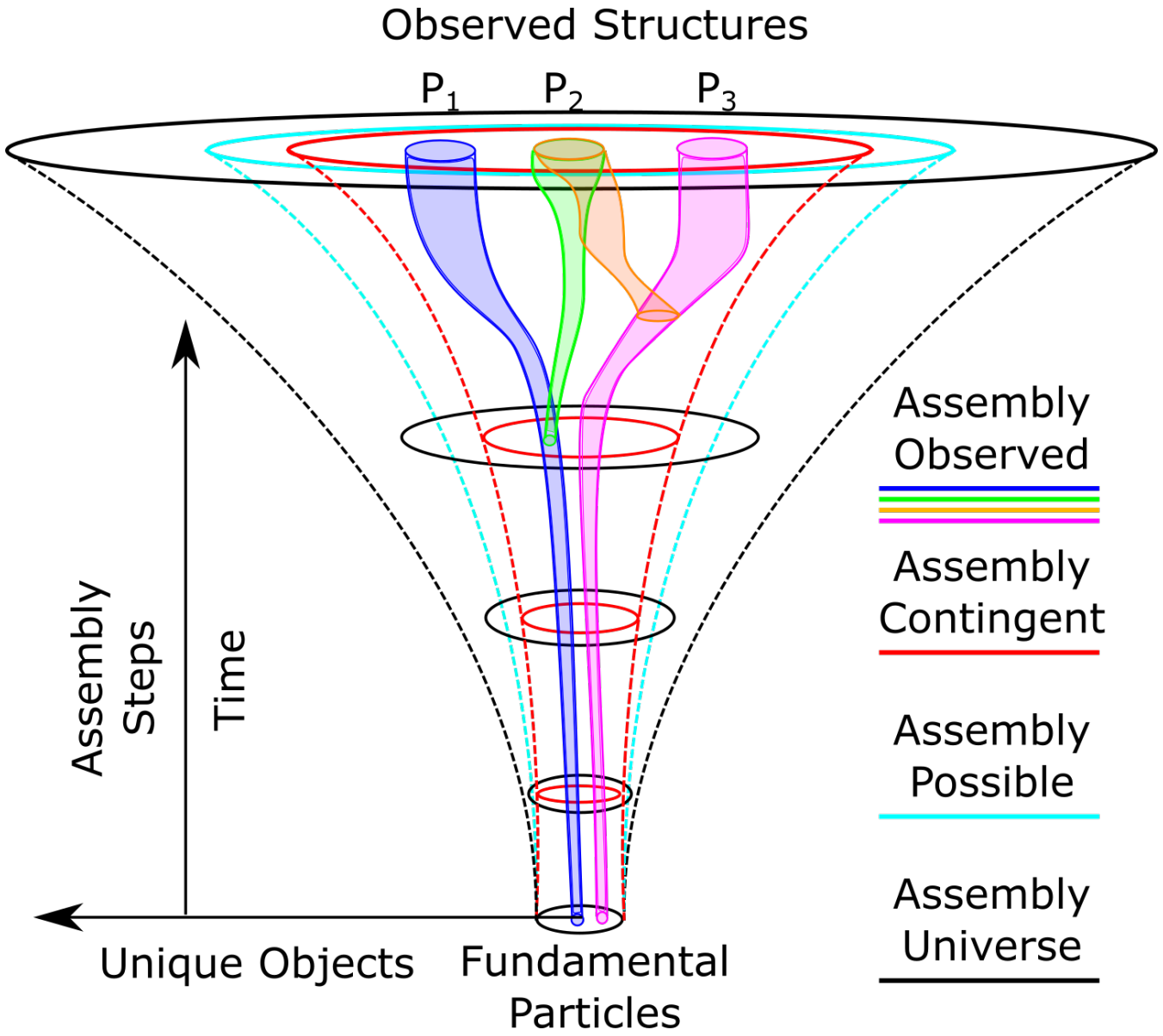

**Figure 4.** Depiction of assembly spaces and their relative sizes where the Assembly Universe, $A_U$ > Assembly Possible $A_P$ > Assembly Contingent, $A_C$ > Assembly Observed, $A_O$.

Possible futures are causally open and undetermined[36,37], but the possible past is causally closed and recorded in existing objects. This implies that the future is predictable only in so far as structure selected in the past can continue to exist, leading to an emergent determinism where features of the present are persistent, deterministic and even predictable because of selection for these features in the past. The assembly space is derived from the deconstruction of observed objects in the present: the units and joining operations are determined via recursive fragmentation of observed objects. We define fragmentation as the process of splitting an object into two parts, e.g., $z \rightarrow \{x, y\}$. By symmetry and invariance of physical law[38], the inverse causal operations of joining the two objects must then also be possible, defining the joining operations and the assembly space, see, e.g. Marshall for molecular assembly space[23,39]. We note the physical circumstances in the environment mediating a fragmenting process or joining process need not be strictly equivalent, for example it may require human-level intelligence to make a highly symmetric DNA nanostructure, but the molecule could be broken by the ordinary abiotic process of phosphate ester hydrolysis. Thus, the symmetry of causation we invoke here does not imply that the environmental constraints or constructors allowing the reverse (fragmentation) and forward (joining) processes are symmetric (see e.g., Marletto and Vedral for an example of similar asymmetry in fundamental physics[40]). Instead, here symmetry and invariance imply that the causation *in* the material (fragmenting/joining) *is* symmetric within the assembly space. One arrives at units when an object produced by a particular mechanism of fragmentation (e.g., by breaking bonds) can be fragmented no further by that same mechanism, defining the boundary of a 'substrate' in AT. The causal symmetry within objects has potentially deep implications given the connection between invariance and symmetry[41], a subject we leave to future work.

To uncover the structure an assembly space, two requirements are necessary. First, one must identify the bounded, distinguishable structures of a substrate, defining its objects. Second, a measurement scheme must be identified that can interrogate how objects can be fragmented recursively to units. Defining elementary units is non-trivial and will depend on the system's energetics and the resolution of the measuring device. For example, for a molecule the bond is the unit scale; however, for peptide chains the unit scale is the amino acid, not the bond. In general, units are defined at the scale where persistent structure becomes combinatorial. From units, paths are re-assembled via causal joins to find the minimum number of steps to arrive at the observed objects.

Following from the definition of objects in AT (see Sharma et al.[22]), 'objects' in the assembly space are strictly those observed to exist in high enough copy number to be measurable[42]. These represent the lineages of causal steps in the assembly space where there is a high copy number at the terminus, $A_O$, see **Figure 4**. All other structure in the assembly space is what we call *'virtual objects'*: their identity is defined based on the existence of objects, e.g., via the process of fragmenting those objects and re-assembling them. Virtual objects are not necessarily independently observable, because there need not exist a causal mechanism that can produce identical, autonomous copies of these structures in high numbers. However, virtual objects along the lineages comprising $A_O$ are causal and therefore as physical as the observed objects.

The definition of an assembly space is predicated on objects that can be deconstructed in a finite sequence of causal steps. In the prevailing conception[26] of reductionist physics, breaking apart physical systems is understood to reveal something of the laws that bind objects into higher-order structure[43]. For example, removing electrons from an atom reveals properties of Coulomb interactions between the negative electrons and the positively charged nucleus. This has led to a standard explanatory interpretation in the form of physical laws that exist outside of objects and somehow dictate their behaviour, leaving open the question of what 'breathes fire' into the equations of physics to yield a universe for them to describe, as articulated by Hawking[44]. In AT, we view the reductionist insight differently, as it reveals a more general property of all objects our universe constructs: the causal constraints from which an object can be assembled are *intrinsic* to that object. There are no external 'laws'. Our view appears consistent with a causal powers interpretation of physical law[45]. This sets the foundation for defining the assembly index as an intrinsic measure of causation, leading to assembly index as an ontological complexity measure (see, e.g., Lloyd[46] for other categories organizing prior approaches to complexity across different fields).

**Sidelight on Possibility and Probability**

To understand the physics underlying assembly theory (AT), it is important to first make a distinction between the concepts of "probable" and "possible". To be *possible* indicates an event can happen, or an object can exist (by existing causal mechanisms as no statement can be made for causation that does not exist). Statements of possibility should *not* be confused with statements of likelihood. It is not meaningful to assign likelihoods in cases where there is (yet) no mechanism for the object in question to exist.

In AT, we do not regard statements of what is probable to encode *any* information about what is possible. Our motivations underlying distinguishing the probable from possible are in contradistinction to other accounts of probability, e.g., the most common being frequentist accounts[47–49], propensity accounts[50], or Bayesian accounts[51,52]. To understand this distinction, consider a statement like *"Object A has a probability $\rho(A)$ to occur"*. Current accounts of probability will treat such a (physical or epistemological) statement as being about $A$, capturing the tendency, or belief about tendency, for $A$ to occur, in many cases conflating probability with possibility. In AT, we disentangle probability from possibility: probabilistic statements about $A$ are not intrinsic to $A$; instead, such statements are about the system and environment that generated $A$. That is, $\rho(A)$, is a feature of a causal mechanism that can assemble $A$ (this can hold for either a physical or epistemological interpretation of probability in AT). Some form of causal mechanism must exist for $A$ to be observed. In short, in AT, we do not regard probability as saying *anything* intrinsic about the object; rather, probabilistic statements are those that encode information about the environment that made the object. Consider the example of ribosomes: such structures are possible everywhere in the universe (since we know these are possible here), but the only planet in the universe where they are probable is Earth. It is also possible that ribosomes will be probable in the future on other worlds, e.g., if Mars becomes inhabited with life from Earth; but no statement can yet be made (as of writing) about the probability of that possibility because it does not yet exist.

Possibility is encoded in the assembly space $A_P$, where the causal possibilities of observed objects indicate others that are possible. Combined with the definition of an object in AT, possibility is open and undetermined, but the expansion of what is possible is dependent on what objects exist and the mechanism of novelty (to be discussed later). Probabilities can be assigned as a phenomenological means to encode the role of the environment of an object in that object's assembly, e.g., assigning probabilities to steps allows talking about the production rate or error-rate of making a specific object with a given accuracy (an example is in how the probability for forming a ribosome should be assigned to the cell, not the ribosome). Assigning probabilities to events or objects is only possible if those events or objects are already selected to exist, and this is not a feature of the object because the mechanism to generate the object exists outside of it.

The assembly space explains how evolution and selection allow 'compactifying' larger volumes of causal possibility within smaller volumes of physical space, increasing the virtualization of objects as they become deeper in assembly space, that is, as 'life' or 'intelligence' carve deeper trajectories into what is possible. A further point on this concept of possibility is the key tension that exists between the physics of an open-ended universe in assembly theory and standard assumptions in analytic modal metaphysical descriptions of contingency and possibility. Modal logic S5 is a widely used framework for metaphysical modality[53] and treats possibility as fixed and eternal. That is, in S5 if something is possible, it is necessarily possible ($\Diamond P \rightarrow \Diamond P$, where $\Diamond$ denotes possibility and denotes necessity). This carries the implication that modal (possibility) space is a static background independent of causal history (e.g. it is Platonic). This view is deeply embedded in Lewis's idea of possible-worlds[54,55].

Assembly theory is in fundamental tension with this picture. The assembly space jointly determines what must occur *and* what can occur given existing constraints, thus, both necessity and possibility are indexed to causal history rather than fixed transcendently. This is consistent with a tradition of intrinsic, dispositional accounts of modality[56,57], and with philosophical arguments that modal structure is itself historically constituted rather than given in advance[58,59]. In assembly theory, there is no transcendent ground of necessity outside what currently exists: what is possible and what is necessary at any moment is a product of what has come to exist, making modality fully physical and dynamic.

We take these distinctions to other epistemological and ontological views to be a philosophically well-motivated feature removing the dependencies of current physical theories on entities that exist outside of the universe being described. It is worth noting that *any* statements of possibility one might choose to make (including logical and nomological ones) are consistent with the view of assembly theory: whereas it is necessary that statements about possibility *must* necessarily be conditioned on existence so we can even make them, it is not true that our ability to make such statements necessarily reflects truths outside of what exists.

**Causation**

An axiom of AT is that objects require a causal sequence of steps to exist. Indeed, objects *are* a causal sequence of steps. Thus, there must be a minimum number of steps as a necessary property for the existence of any object. This physical property of each distinguishable object type, $i$, is quantified by assembly index, $a_i$. Assembly index is ontological and is an *intensive* property, subject to measurement, independent of formation history, system size, or measurement method (see section on measurement). We now define it more formally[34].

**Definition 2 - Assembly Path and Assembly Index**. An *assembly path* is a finite sequence of causal joins, $\{(x_1, y_1, z_1), \dots, (x_n, y_n, z_n)\}$ that satisfy: $\forall i \in \{1,2,\dots,n\}$, $x_i, y_i \in U \cup \{z_1, z_2, \dots, z_{i-1}\}$ and $J(x_i, y_i, z_i) = 1$. That is, for any object $z \in \Omega$, an assembly pathway for $z$ is defined as a finite directed acyclic graph whose leaves are labelled by elements of $U$, whose internal nodes represent causal join operations $J$, and whose join operations in the pathway ultimately produce a single final object $z$. For any object, $z \subset \Omega$, the *assembly index* of that object, denoted $a_z$, is the number of causal joins along a shortest assembly path that has $z$ as its final product. If $\{(x_1, y_1, z_1), \ldots, (x_n, y_n, z_n)\}$ is a shortest assembly path for $z$, then $a_z = n$. For a canonical choice of primitives, composition rules, and object equivalence, the assembly index is a well-defined property of the object.

While it is simple to mathematically write out an assembly pathway as a labelled sequence, doing this in a way where the labels are accurate to the causal structure of the assembly space is non-trivial. Joining operations in an assembly space are substrate-specific and must encode causation as it operates in that substrate. For example, in molecular assembly space, hydrogen atoms are typically not treated as explicit assembly primitives, because including them adds substantial combinatorial complexity without increasing discriminatory power in practical measurement nor in the calculation of molecular assembly index[34]. However, for atmospheric assembly space they are critical due to the hydrogen bonding between gases[33]. The assembly space of solid state materials has yet a different structure[31].

The formalization of assembly index captures:

(1) ***Causation***. Assembly index counts the minimum number of causes (joining operations) necessary to produce the object.

(2) ***Contingency***. Only structures constructed along an assembly path are available for recursive reuse to construct subsequent objects.

(3) ***Invariance***. Assembly index does not depend on the identity of causal joins or their precise ordering, only the shortest countable number to assemble an object, which is a path-independent feature of the object.

The assembly space introduces a material and intrinsic concept of causation, because it encodes the direct influence of one component on another's possible existence, see **Figure 5**.

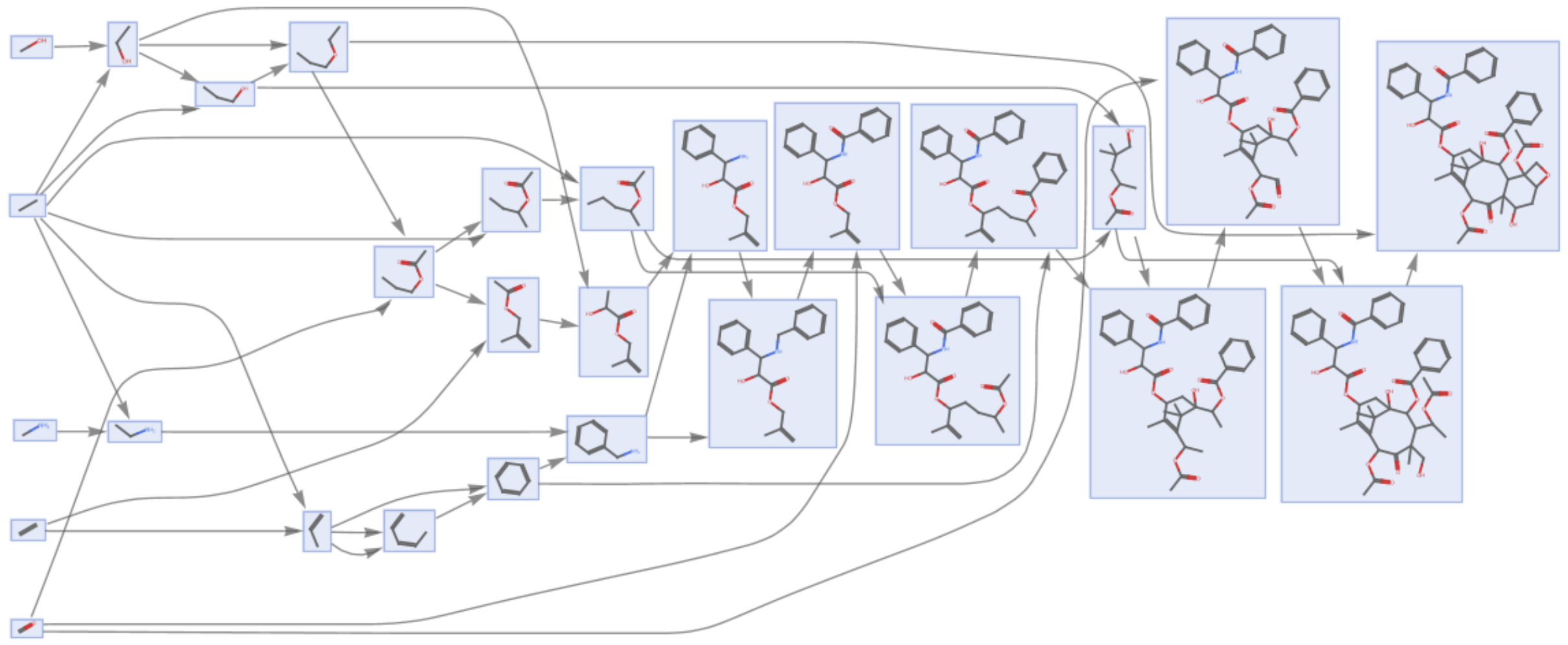

**Figure 5**: Assembly pathways reveal the materiality of causation, where each object within an assembly space exists as a sequence of causal joins on a minimum path, and the object is the high copy number structure at the terminus. Shown is a minimum assembly path of Taxol, which has 23 steps.

AT's material definition of causation dissolves issues inherent to counterfactual definitions of cause[60,61], such as the need to iterate over interventions that cannot be fully specified or computed. The restriction of steps along an assembly path to joining operations implicates these as fundamental units of causation, and yields a precisely defined, and metrological concept of 'cause': causes are structures that appear on a minimal assembly path for an observed object. Every structure on an assembly path is a causal join of two prior structures, and these are not *a priori* labelled: the precise identity as virtual structures need not be known or knowable.

A causal join must exist before its product can be used, imposing a measurable partial ordering on the existence of objects with testable consequences. This yields an ontology where objects include causal extent as a material attribute. Complex objects of the kind we interact with in biology and technology are effectively hyper-objects, as their assembly space, or lineage, is large in its causal depth.

It is not trivial to determine what set of circumstances allows one object to exist over another[62] and most of the causation generating the objects of our everyday experience is so deep in time as to be intractable to scientific study. In AT, we recognize it is not possible to precisely define nor measure the presence of very large lineages of causation direction. It is however always possible to measure the shadow of a long lineage in the object(s) it generates (see also discussion on the connection between AT's concept of cause and Derrida's deconstructivist concept of trace, as explored by Mastrogiovanni[63]).

In physics, what constitutes "matter" and its properties is defined by what can be measured and described via abstractions that allow defining mathematical relationships between measurements. Attempts to formalize life have frequently focused on concepts of information[64], which are context dependent, statistical and/or not standardized across measurement devices. AT is the first to define the physics of information as a property of matter and provide a physical space for 'information' by recognizing causal possibilities as a physical space, the assembly space. This underlies AT's unique definition of material causation. It differs from prior notions of cause in important ways, for example by unifying the Aristotelian notion of a formal cause (the informational plan) as one and the same as his notion of material cause (the stuff something is made of)[65]. AT introduces an ontology where hierarchy is defined in terms of the possibility to exist, radically departing from notions of hierarchy defined by spatial or energetic scales found in other physical theories (see Ardoline[66]).

The concept of 'physical space' in modern physics is taken to be four-dimensional, consisting of three space dimensions and one dimension associated to measuring the simultaneity of events: objects are regarded as existing within this four-dimensional spacetime and can be assigned a position relative to other objects as constituted by measurements of length and proper time. Likewise, the metrology of assembly index establishes it as a physical observable for molecules. Assembly space thus provides a structure for causal possibility: objects exist within this high-dimensional space and can be assigned a position defined relative to other objects (and units) as constituted via the causal joins that link them. In assembly theory, objects do not just move through space and time, but they also move through a space of causal possibilities that is as real.

**Contingency**

Observing *any* object in discrete copies provides direct evidence that the universe contains a causal mechanism to produce (and reproduce) that same object. Copy number is evidence of stored memory. This is distinct from frequentist approaches to probability, where frequency is used as evidence that something is likely[67]. Copy number allows formalizing of contingency along causal chains within assembly space by capturing dependency on prior selective steps.

**Definition 3 - Copy number**. Copy number, $n_i$, is the countable number of each distinguishable object type, $i$, (defined up to limits of measurement).

In AT, copy number is *not* capturing the *a priori* probability of the object occurring, which is, in general, not definable (see Section on Probability and Possibility). Instead, copy number is an indication of contingency in the causal chain that allows the object to exist at all. This is sufficiently important that it is a principle of AT:

**Copy Number (CN) Principle**. Countable copies of a distinguishable object type provide direct evidence of contingency in a causal chain mediating the object's existence.

The copy number (CN) principle is inclusive of mechanisms that are spontaneous. Processes that can occur everywhere (spontaneously) are often captured by what we call "universal laws" that aim to explain such ubiquitous regularities. But the CN principle is broader than universal laws in physics, because it also applies to all non-universal causal regularities, inclusive of high assembly index objects and not just the low assembly index ones. Some mechanisms of causation (indeed most if we consider the full space of possibilities) are contingent and *not* spontaneous, and they are therefore not found everywhere in the universe. Ribosomes provide evidence of one such local causal regularity or 'law-like' process that may only exist on Earth: the constructive context known as a biological cell. Ribosomes have persisted in high copy numbers on this planet for billions of years[68] because there is a localized physical system (the cell) that allows their reliable, reproducible construction. The copy number of ribosomes on Earth is not evidence of the universal probability of their formation, but instead is evidence of a contingent causal chain persisting on Earth.

**Measurement**

Although it is now common knowledge as a statement of scientific fact that water boils at 100°C, this statement is not true of the world. Water boils at slightly different temperatures if it is in a glass container versus a metal one, if it is aerated or not, depending on the altitude you are at, etc. The standardized temperature scale came about because of the invention of a theoretical abstraction, the concept of absolute zero, originally defined as a fundamental limit to the degree of coldness possible[69]. By reference to this theoretical limit, all other (hotter) objects can have a standardized temperature defined precisely. This scale works even if absolute zero is not a physically attainable temperature.

What if there could be a similar standardization, not for a temperature scale, but for the measurement of causation? The assembly index is structured to provide such a standardization, such that each object's assembly index, $a_i$, and the substrate-specific and measurement specific $a_M$, can be determined experimentally. We will describe the metrology of AT as developed for molecules as a special first case of application of the metrology of assembly theory.

For molecules, measurements have been done using techniques where the number of parts within a molecule can be determined using standard laboratory methods, like infrared (IR) spectroscopy, nuclear magnetic resonance (NMR) spectroscopy, and mass spectrometry (MS)[30]. These techniques rely on different types of experimental data, yet each have been found to give evidence for the direct measurement of molecular assembly index.

Assembly theory was first developed by one of us (Cronin) via thought experiments about measuring causation in molecules using mass spectrometry. In a chemistry lab, mass spectrometry is a common technique for identifying molecules and measuring their properties[70]. This works by shattering molecules into fragments, giving characteristic patterns for specific molecules based on the charge-to-mass ratio of the fragments produced. Thought experiments leading to the development of the assembly index were inspired by this fragmentation process, and how one could reassemble a molecule in successive steps by sticking the fragments back together, ensuring any redundancies are removed to guarantee the smallest number of joining operations[23,29]. If some fragments are identical and can be reused, fewer total steps are necessary to reassemble the original molecule. This is indicative of the inherent, physical hierarchy in how to count constructive complexity that came to underlie the assembly index as a measure of intrinsic causation.

This fragmentation process in a mass spectrometer mirrors the formalism of assembly space and the assembly index, so it is tempting to say that assembly index is just a correlate of the mass spectrometry fragmentation, rather than intrinsic to the molecule. However, this is demonstrably not the case. For a given molecule, the complete quantum state (or equivalently its ground-state) contains *all* information about bonding, connectivity, symmetry, and repetition. This can be 'read-out' by the spectroscopic techniques of mass spectrometry, NMR, and Infrared. And, any of these techniques can be implemented to determine assembly index because the assembly index is a functional of the quantum state of the molecule, which measures the minimum generative description length of the molecular graph under physically allowed composition rules. Before AT, this functional was simply not identified, formalized, or measured. What makes assembly index new is not that it adds information to quantum mechanics, but that it selects a particular invariant that had no prior analogue in physics. The assembly index is global property of an object, not local (unlike energy, charge density, or force constants), it is non-extensive (copying a module does not increase it linearly), and it is algorithmic, not additive. The assembly index is observer-independent once the construction rules are fixed. Finally, it is experimentally accessible through lossy projections (IR, MS, NMR) without full structure reconstruction[30]. This is why IR, MS, and NMR can all be used to measure the assembly index, using radically different physical mechanisms and instrumentation. These techniques are not measuring the same thing directly; they are however all projecting the quantum state through different symmetry, and redundancy preserving, channels. The same scalar emerges robustly across these orthogonal probes providing the strongest current evidence that assembly index captures something intrinsic rather than an artefact or some kind of heuristic. This provides a newly identified intrinsic physical invariant of molecular structures, consistent with quantum mechanics but not recognized, nor experimentally operationalized until the advent of AT.

Thus, via the assembly index, the concept of 'causation' in AT becomes a feature like charge, spin, or other quantum numbers, or the speed of light $c$. It does not depend on the object's environment, reference frame, or measuring device. Although this is a strong claim, it has been rigorously tested[23,30]. If the assembly index were not intrinsic, it would depend strongly on representation, spectroscopy choice, or observer-dependent assumptions. The empirical fact that it survives IR, MS, and NMR, all with different physics, different noise, and different symmetries, suggests invariance[30].

This invariance is critical to metrological standardization of the constructed complexity of molecules, which is a necessary step towards experimental confirmation of living threshold for biosignature science. To confirm a threshold, measurements must be taken for objects derived from living and non-living samples, with an understanding that the threshold value is system size dependent, $a_M(N_T)$. For a laboratory mass spectrometer, the copy number of identical molecules sets a detection limit of $M \cong 10{,}000$ because this is the lower limit of detection, see **Figure 6**. As in **Figure 3**, we set $b = 12$ to be consistent with experimental measurements in Marshall et al.[29], which set the limits of detection of abiotic molecular complexity at $a_M{\sim}13$ (vertical, black dashed line Figure 6) in millimolar samples. In Figure 3., there is no measurement cut-off, but here we show how the tail of the distribution will be below the measurement cut-off for most experiments, rendering the two distinct regimes (spontaneous or abiotic, versus selected or life) more obviously distinguishable by an abrupt discontinuity (see e.g. Marshall et al.[29] where the cutoff at $a_M{\sim}13$ is abrupt and for $a_i > 15$ molecules are found only as derivatives of biology). For the simple model presented, a value of $a_M{\sim}16$ is expected for experiments on abiotic systems with $10^3$ mol of material, a prediction which can be refined with more detailed analyses of the structure of molecular assembly space and its branching, alongside instrument resolution considerations.

We show this detection limit set against the abiotic assembly copy number distribution in **Figure 6**, showing how determination of a living threshold in assembly space depends on both system size and instrument resolution. The existence of a threshold has been demonstrated experimentally in Marshall et al.[23]

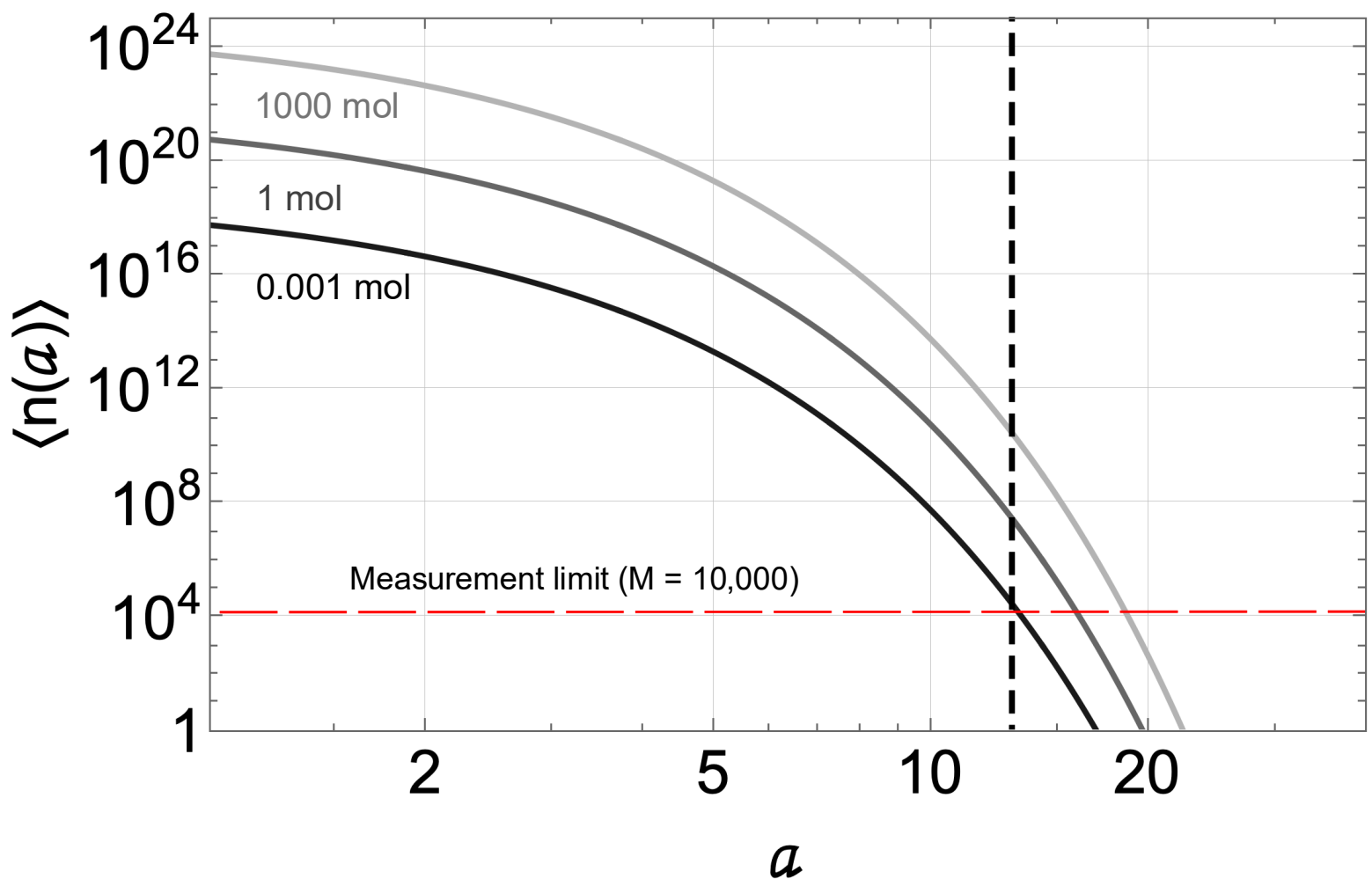

**Figure 6:** The epistemological limit set by a measurement device determines the maximum assembly index object expected to be observed in the absence of selection. Shown is the measurement resolution for a mass spectrometer, where a minimum of M = 10,000 identical molecular objects must be present (horizontal red dashed line), relative to the abiotic copy number distribution for laboratory scale samples ranging from 1/1000th of a mole to 1,000 mole.

**Selection**

A generalized theory of selection beyond what is described in biology is necessary if we are to identify how selection gives rise to biology. In developing AT, we consider selection as a generalized phenomenon, where combinatorial possibilities are restricted to yield the set of persistent objects.

We are now able to define the *Assembly (A)* of a system, which quantifies the total amount of selection among causal possibilities required to generate and maintain all objects of a system. It combines both the material causation of objects (assembly index) and the evidence of persistent contingent causal mechanisms (copy number), allowing a formalization of non-living and living (or intelligent) objects. Living objects take the form of an assembly contingent space, which represents the existing space of objects constructed by selection.

We call the configuration of causal possibilities intrinsic to a collection of objects an *assemblage*, where the total causation is quantified by its Assembly as:

**(3)**

$$A = \Omega \frac{1}{N_T} \sum_{i:n_i>1} n_i \, e^{a_i}$$

where $n_i$ is the copy number and $a_i$ is the assembly index of the $i^{th}$ distinguishable object type, $N_T$ is the total number of objects and $\Omega$ is the Assembly constant (for molecules this is in the units kJ $mol^{-1}$). This equation captures how producing few copies of a complex object requires much more selection and contingency along causal lineages, than producing many copies of a simpler object.

A configuration in standard physics is formalized to capture the current position and momentum of a system. This should be considered only as the current instance of that configuration's assemblage: the assemblage includes the accumulated causal possibilities, recursively assembled to generate the configuration. The assemblage is a causally deep representation of matter.

We take as an axiom that any object with assembly index $a_i > 1$ cannot be realized without the existence of objects of lower assembly index along at least one minimal assembly pathway; that is, every object observed with assembly index $a_i > 1$ exists both in the physical objects we count and interact with, and as a virtual space of finite support that encodes the underlying causation necessary to its existence. That is, each object observed implies a history of an accumulated hierarchy of objects causally prior to its existence. The quantity $A$ makes this implication explicit, defined by introducing the *virtual copy number* to account for copy numbers along causally contingent chains leading to the observed objects. To derive this, we assume (virtual) copy number per observed object grows continuously over time in the past:

$$v(t) = (1 + h)^t$$

Where $h$ is the growth rate and $t$ is a continuous time over which objects were created, leading to the final object. We define the growth rate as

$$\Delta t = h = 1/P$$

where $P$ is the number of possibilities from which a single virtual object is selected. That is, the growth rate of virtual copies in the past is defined as the minimum unit of time to form a single (virtual) object from $P$ possible objects. The observable finite number of causal steps is fixed by the assembly index of the observed object:

$$a = t \cdot \Delta t = t/P$$

We note that this equation provides the connection between causal time (measured in assembly steps, $a$) and continuous time (as measured by clocks, $t$), where the proportionality is driven by the size of the possibility space undergoing selection ($P$). Rearranging yields the total coordinate time for virtual object formation embedded within an object:

$$t = a \cdot P$$

With this, the discrete causal time virtual copy number equation becomes

$$v(a) = (1 + 1/P)^{a \cdot P}$$

A key conjecture of assembly theory is that selection operates in a space of continuous undefined possibility, where the future is open and causally undetermined and objects become discrete only in the present. This renders the past computable but the future uncomputable.  To introduce selection against the continuum of possibility, e.g., against an unresolved virtual background of causal possibilities from which new objects come to exist for the first time (related to time itself as continuous and creative in the absence of causation) we take $P \rightarrow \infty$, recovering the form:

$$v(a) = e^{a}$$

It should be noted that the exponential form in the assembly equation is derived from the assumption that selection is what discretizes a continuum of possibilities into objects; thus, the mechanism of selection is embedded in the structure of the virtual copy number. We note that $v(a)$ is determined for each observed object; therefore, we must multiply by $n_a$, that is, we account for the copy number of observed objects by multiplying the copy number of each observed object by its virtual copy number burden. Summing over an entire assemblage yields the total virtual copy number of the population:

$$v_{\text{tot}} = \sum_{a} v(a)\,.$$

And finally, the assembly, $A$, of a set of objects is the total virtual copy number per observed object,

$$A = \Omega \frac{v_{\text{tot}}}{N_T} = \Omega \frac{1}{N_T} \sum_{i:n_i>1} n_i\, e^{a_i}$$

Here, $\Omega$ is a constant measured in kJ mol$^{-1}$. $A$ represents a causal density per observed object. It is measurable because both $n_i$ and $a_i$ can be experimentally determined. The restriction to object types with $n_i > 1$ implements the copy-number principle: repeated occurrence is necessary to evidence a reproducible, selected causal generative process rather than a fluctuating contingency. $A$ quantifies how, for each observed object, there are infinitely many infinitesimal selection steps embedded in the finite interval of $a$ metrologically accessible steps, which characterize the causation in the selected assembly space within the object.

This formulation allows us to quantify when systems undergo transitions from abiotic chemistry to constructed complexity ('life'). As a system accumulates more complex objects in higher copy numbers, its $A$ increases. Systems above a critical threshold in $A$ mark a transition in assembly space, and objects with $a_i > a_M$ require contingent lineages of objects, and memory, as necessary for their persistence. Such objects are causally deep in the assembly space and are expected to exhibit properties commonly associated with living and intelligent systems, including persistent complex structure, reliable reproduction, and potential for agential and open-ended behaviour. These are features which we regard as arising because of their deep virtualization in the space of causal possibilities.

**Open-Ended Evolution, Novelty and the Mechanism of Selection**

Assembly operates on discrete objects whose identities and assembly indices are defined retrospectively, while interactions between possible objects occur against a continuous or unresolved future that is not discretised in advance before selection fixes causal mechanisms. Novelty emerges when such interactions produce outcomes that do not map onto existing object types. Most such outcomes are transient and leave no causal trace. Persistence enables repeated interaction with the same unresolved region of possibility space, mediating its discretisation to new physical object types. Once discretised, an object becomes part of extant causal memory and constrains future interactions. Assembly index records the depth of this accumulated causal memory.

The interaction of an undetermined future with a causal present provides a mechanism for open-ended evolution, where open-endedness emerges from an asymmetry between discrete causal memory and an under-resolved environment. Open-ended novelty therefore arises not from randomness, but from delayed discretisation of continuous interactions filtered by persistence and reuse.

Assembly theory treats a population of objects as a discrete causal memory embedded in an undetermined space. Each object $i$ is characterised retrospectively by an assembly index $a_i \in \mathbb{N}$, the minimum number of joining operations required to construct it, and a time-dependent copy number $n_i(t)$, with $N_T(t) = \sum_i n_i(t)$. Assembly index encodes causal depth but does not predict which objects will appear.

Novelty arises when interactions between the current assemblage and its under-resolved environment generate outcomes not previously present; these outcomes are only discretised into new object types if they exist and can recur, because only then can their identity be indexed and measured.

Depth and persistence are connected via virtual copy number. One copy of an object at assembly index $a$ implies an exponentially growing causal structure supporting the existence of the object, so we associate with each object type, $i$, a virtual copy number $v_i(t) = e^{a_i} n_i(t)$. The assembly becomes time-dependent in the normalised virtual burden carried by reproducible objects,

**(4)**

$$A(t) = \Omega \sum_{i:\, n_i(t)>1} e^{a_i} \frac{n_i(t)}{N_T(t)} = \frac{1}{N_T(t)} \sum_{i:\, n_i(t)>1} v_i(t),$$

which quantifies how much causation is instantiated per observed object.

A discovered object may exist transiently with $n_i(t) > 0$ yet remain causally sterile; it becomes causally effective only if it persists long enough to be used to propagate formation of structure. This means the object's persistence time $\tau_i$ must be long enough for the copy number to increase, e.g. such that $n_i(t) \geq 2$. Causation at assembly index $a_i$ requires this persistence time to exceed the characteristic waiting time, $\tau_C$, for reuse in the combinatorially expanding assembly space such that $\tau_i \gtrsim \tau_{a \to a+1}$. When this condition holds, the virtual copy number causal burden per object $e^{a_i}$ implies the causation is trapped once and replicated through reuse; when the condition fails, the burden must be repaid upon each rediscovery, and the object cannot contribute to propagating causation nor to sustainable growth in assembly index.

This leads to an expansion and contraction of the assembly space. When new objects are constructed, they expand the assembly space itself, enabling construction of even more complex objects. This can yield a "ratchet" effect where each new object opens new possibilities, and the extinction of objects closes off possibilities. In expansion phases, the threshold recedes, allowing spontaneous combination to make more complex things than in the absence of a deep lineage. In contraction phases, the threshold drops back, returning to the abiotic value, $a_M$, in the absence of any deep lineages. Thus, the assembly space is not static but dynamic, becoming larger as new causal mechanisms emerge and smaller as objects die out. The new causal mechanisms cannot be predicted *a priori,* and mechanisms are only predictable so long as they are under selection and can persist.

AT provides a testable mechanism for open-endedness following from delayed discretisation at the interface between discrete memory and continuous possibility. Interactions continually sample an outcome space that cannot be enumerated in advance; most outcomes vanish without leaving a causal trace in $A(t)$. Those that persist are stabilised, labelled as new object types (with $a_i$ defined only after the fact), and incorporated into the assemblage, increasing the mass of reproducible high-$a$ objects and allowing the high-end tail of the assembly-index distribution to ratchet upward over time only under certain selective circumstances (e.g. as governed by the persistence time of objects).

Selection acts forward on persistence and copy number, while causation accumulates backward through assembly index, and their coupling is made explicit by the joint dependence of $A(t)$ on $a_i$ and $n_i(t)$. In other words, any genuine, novel, open-ended process does not have objects that can be defined before the universe constructs them, captured in the ontology of AT in how the assembly space can only be determined relative to objects that exist in the present. There is no definable, global assembly space that exists outside the objects we observe, enabling a physics describing a universe that is self-constructing.

Deterministic dynamics and predictability are expected as emergent properties of the assembly space: more objects become predictable as objects grow deeper in the assembly space because the space itself will encode more causal mechanisms. Many physical processes can be simulated algorithmically after the fact, but open-ended evolutionary processes cannot be fully characterised as closed computations over fixed symbols and rules. The reason is structural because the identities of relevant objects and operations are not fixed in advance but emerge retrospectively through persistence and reuse. The effective state space therefore expands in a way that cannot be enumerated *a priori*. Any finite history can be simulated retrospectively, but no single closed algorithm can predict in advance which new objects will arise, persist, or become causally significant. This reflects delayed discretisation and state-space expansion and contraction, determined by what objects persist.

**Entropy and Assembly**

To understand $A$ as a physical quantity, we can relate it to more familiar concepts like entropy. From our discussion on measurement, it should already be apparent that assembly index is unlike entropy, as it is not an ensemble quantity. Assembly index is more like a topological or algorithmic invariant. Once the construction graphs are fixed, it is a property of the individual object (and therefore, for molecules, of the underlying quantum state that determines that structure). Assembly, by contrast, does play an analogous role to entropy, but for structured objects under construction, not for ensembles under thermal mixing.

Consider a toy example of two simple molecules with identical molecular formulae, $C_{11}H_{24}$, but with different structures (they are structural isomers); 4-propanyloctane and 2,3-dimethylnonane. Assembly index can characterise the amount of causal symmetry in a molecule (how many steps are repeated), see **Figure 7**.

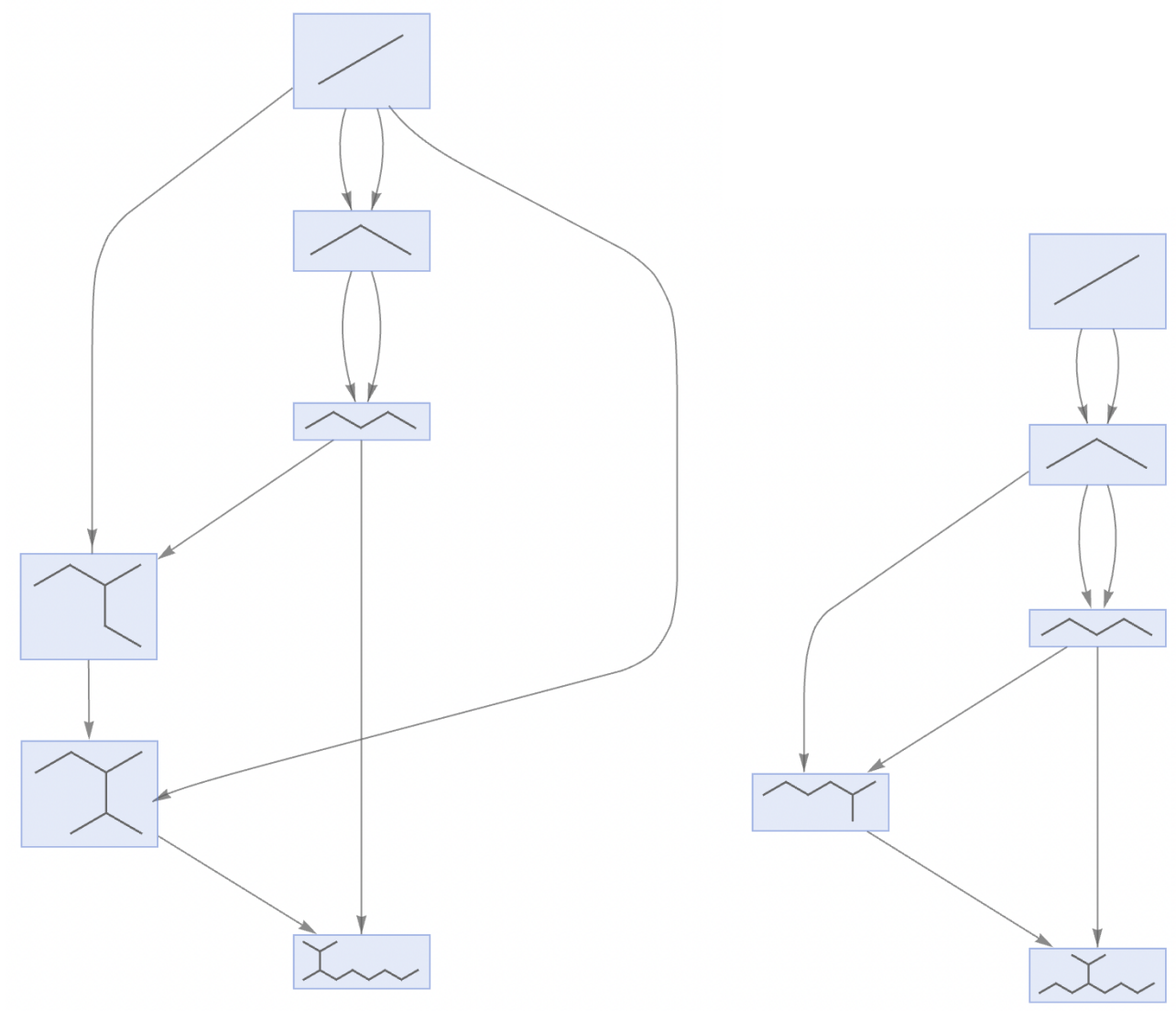

**Figure 7:** A minimum assembly path for 2,3-dimethylnonane (left), 4-propanyloctane (right). Both molecules have the same molecular formula, $C_{11}H_{24}$, and the same Shannon entropy for their SMILES representations; (CC(C(C)C)CCCCCC and CCCC(C(C)C)CCCC of 3.8 respectively. They differ in their physical properties including assembly index, with 2,3-dimethylnonane being $a = 5$ and 4-propanyloctane being $a = 4$ respectively, and heat of formation with 2,3-dimethylnonane being -255.7 and 4-propanyloctane being -179.4 kJ $mol^{-1}$, respectively.

Molecules with different symmetries are also energetically distinguishable. The molecules 2,3-dimethylnonane and 4-propanyloctane have the same molecular formula, but differ in both their assembly index, which are $a = 5$ and $a = 4$ respectively, and in their heat of formation, which are -255.7 and -179.4 kJ $mol^{-1}$ respectively. The less symmetrical molecule has the higher assembly index and the higher heat of formation.

Next, consider the following thought experiment in which we assume the production of the molecule Taxol increases fitness, and hence its production increases with the growth and persistence of the Yew tree. A Taxol Demon represents the construction machinery that assembles the Taxol molecules using the gases as reagents. The contribution to assembly, $A$, from Taxol will increase in a sigmoidal fashion, see **Figure 8**, representing the constructed order in the molecules, mirroring the decrease in entropy.

Recent advances in statistical physics have emphasized the importance of path-dependence in non-equilibrium systems[71,72]. Assembly also formalizes a concept of path-dependency, which does not abstractly assume all possible paths are physically real, nor does it assume we need to experimentally select a path directly for it to be real (e.g. we do not assume context dependency in the measurement of paths). The space of all paths cannot be exhaustively sampled (nor computed) in any real complex physical system of interest, suggesting it does not make physical sense to assign probabilities over these spaces: they are too large for even the entire universe to ever sample[73]. One might attempt to sample based on a set of identified constraints, but a deeper challenge emerges, because the constraints themselves must be selected from the space. AT replaces the unphysical assumption of state spaces that cannot be sampled with a precise, finite definition of an expanding assembly space capturing object-dependent microphysical causation and contingency, which could provide an alternative foundation for path-dependency in non-equilibrium systems based on causal principles.

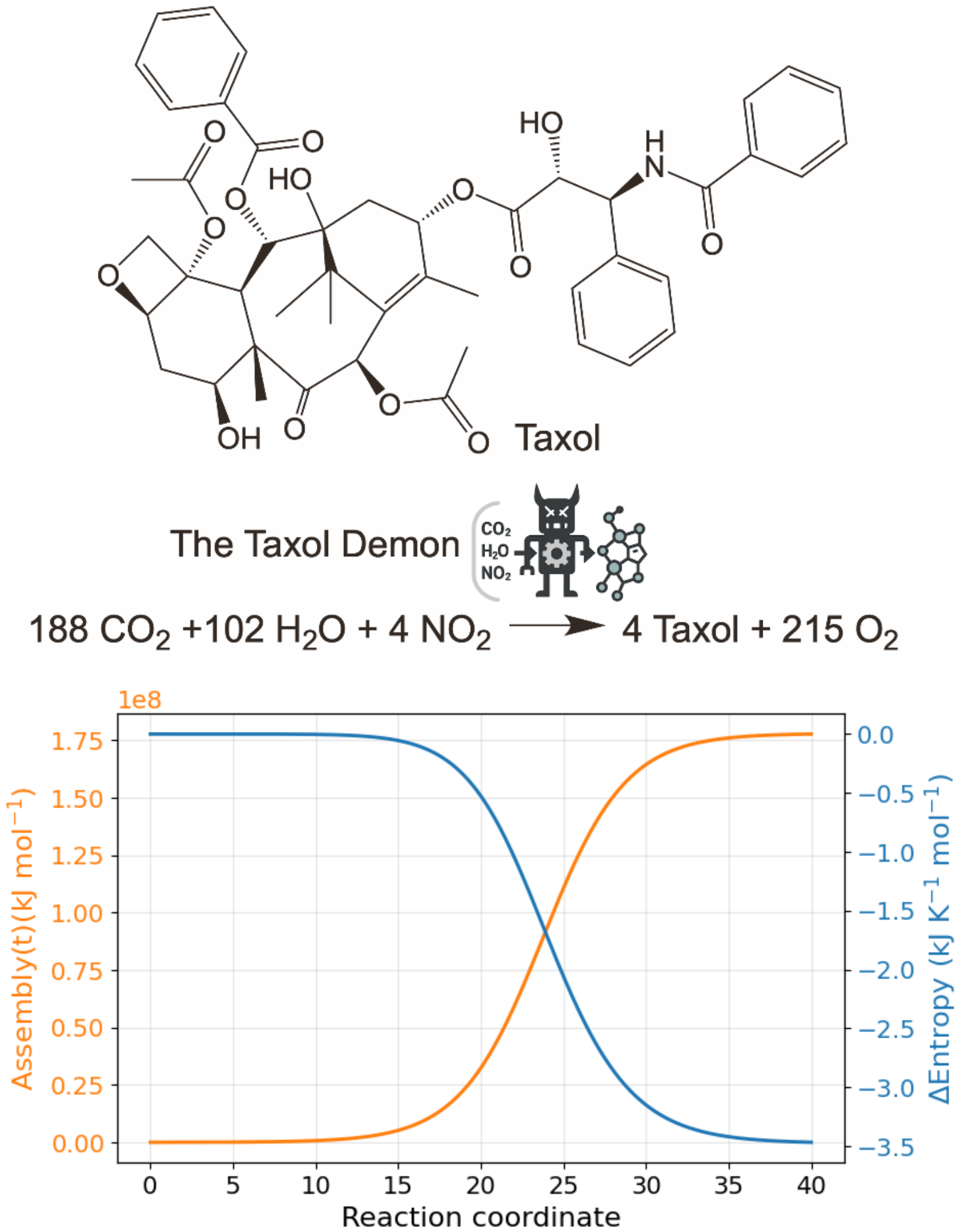

**Figure 8.** Top: A depiction of the process to construct Taxol from what would be its combustion products in the presence of oxygen, and the hypothetical Taxol Demon. Bottom: The increase in assembly (*A*) in Taxol for 1 mole of the natural product from the component gases building blocks, $CO_2$, $H_2O$, $NO_2$ for this thought experiment. The entropy change of the system is depicted. With the maximum increase in assembly reaches 1.75 $\times 10^8$ x $A_k$ in kJ $mol^{-1}$ (the Assembly constant, $\Omega$, is set to 1 in this model) the entropy of formation is -3.8 kJ $K^{-1}$ $mol^{-1}$. The entropy change was calculated using standard molar entropies of the gases and estimating Taxol's molar entropy to be 1000 J $mol^{-1}$ $K^{-1}$.

### Causal Transitions and Formalizing "Life"

We are now able to discuss the nature of causal transitions within AT, and how these allow the formalization of "life" characterized by persistent copies of objects in abiotically inaccessible regions of the assembly space. Here we identify causal transitions as those where a combinatorially explosive system becomes restricted, allowing propagation of structure into deeper regions of assembly space. Causal transitions underlie the persistence of objects beyond the living threshold. As we have indicated, the assembly index for each object is intensive to that object, as it does not depend on the amount of sample. But $A$, and $a_M$ are extensive, where both depend on the amount of material through the number of distinguishable objects $N_T$. The abiotic threshold $a_M$ defines a critical surface in the space of possibilities, where any observed object with a larger assembly index than this, and which persists in multiple copies, can only be produced via a selective mechanism. Because $a_M$ is extensive, the value indicative of 'life' will be system size dependent.

For each substrate-dependent assembly space, it should be possible to derive the conditions at the *assembly threshold*, $a_M$, marking the upper limit of the complexity of non-living systems in that substrate, which will be system-size and substrate dependent. In general, our simple model presented earlier will underestimate the size of exhaustive search because we assumed a fixed branching to the possibility space, when we should expect the number of possibilities to increase with increasing assembly index. Nonetheless, we here use the same simple model to also motivate some expectations about the behaviour of $A$. Substituting equations (1) and (2) into equation (3) yields an effective model for $A$ in the absence of selection (see **Appendix B**):

**(5)**

$$A_{abiotic} \propto N_T^{\frac{1+\ln b}{\ln(1+b)}-1}$$

To arrive at this form, the summation in Eq 3 is done only up to $a_M$ (because for $a > a_M$, in the non-living regime the terms are unobservable or zero). This is shown in **Figure 9**.

In prebiotic chemistry, molecules with $a > 13$ are sometimes experimentally produced but these have not been observed abiotically[74]. To apply assembly theory to prebiotic chemistry it is important to note ***all*** prebiotic chemistry experiments include some degree of selection, which comes in many forms ranging from using purified reagents, to controlling process conditions, to tightly controlling yield that are rarely reported, as well as introducing reagents with a high assembly index[75]. It is an open question how to best parameterize and quantify these selective constraints[76].

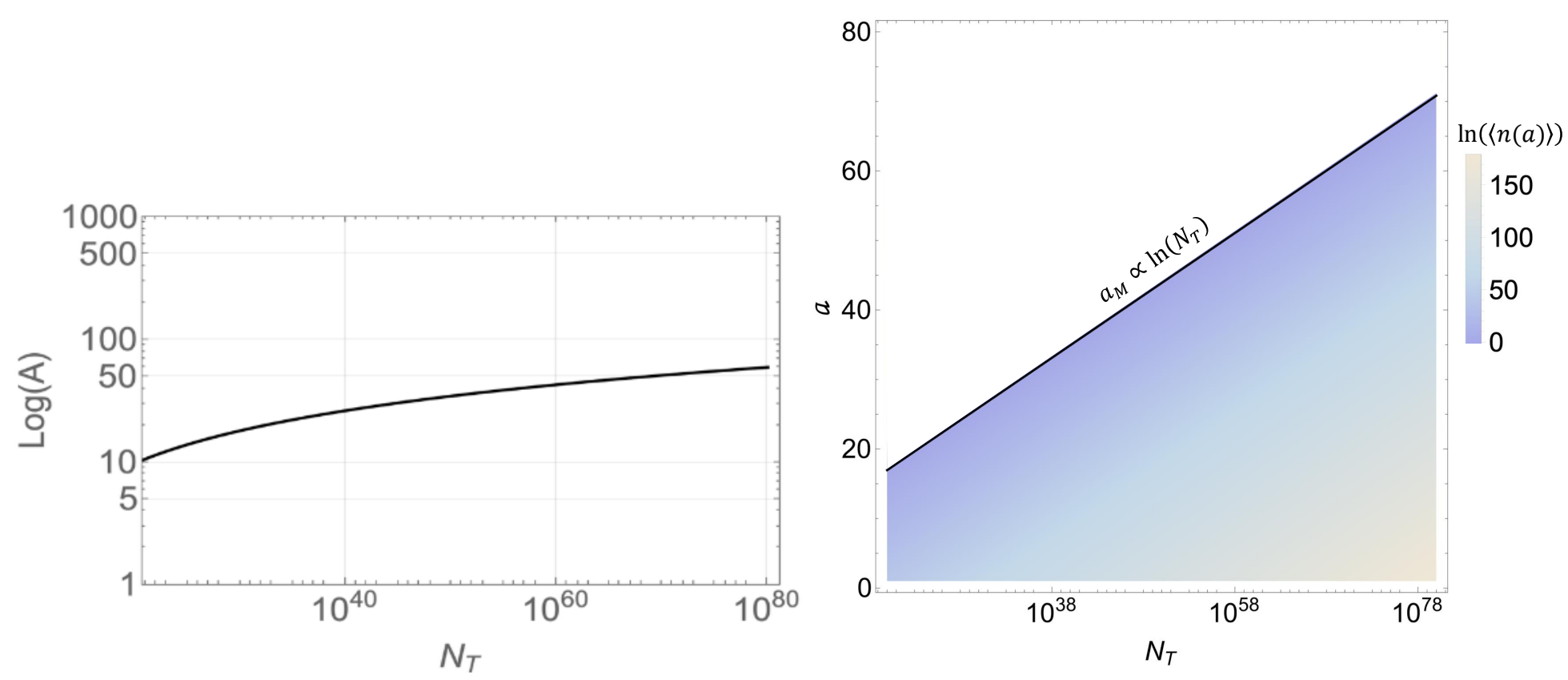

**Figure 9:** Top: The abiotic assembly $A$, consistent with a uniform copy number distribution over objects and a threshold $a_M$, decreases with the number of objects $N_T$. Shown is the distribution as a function of $N_T$ and the assembly for ATP at concentrations found within a typical bacterial cell. Bottom: Abiotic copy number $n(a)$ as a function of $a$ (y-axis) and $N_T$ (x-axis) for fixed branching ($b = 12$, as above), with the line for $a_M$ showing the division between objects expected to exist and those which are too assembled to exist in the absence of selection.

To be clear on the distinction, we use *'abiotic'* in the strictest definition of the chemistry that preceded life, which had no human influence or design[77,78] and included only unstructured (or random) biases in reaction rates, which would be insufficient to constrain the combinatorial explosion of branching in chemical space to drive production of molecules beyond the threshold. We use the term *'prebiotic'* to be used to denote specific selective circumstances directed towards producing the molecules found in biology on Earth, which defines most current experimental efforts in prebiotic chemistry. That is, 'prebiotic' describes those cases where post-selection on extant biology is used to constrain experimental design on its origin (e.g. justifying the prefix "pre" denoting 'before' as in 'predict', whereas the prefix 'a' denotes 'not/without' as in 'asymmetrical'). Prebiotic chemistry experiments represent significant selective constraints, placing nearly all experiments in the origin of life chemistry in the regime of causal structure associated with life[64], which is manifest in precise control by intelligent agents in the design and execution of prebiotic chemical experiments. Our hope is that AT will open new paths towards proper bookkeeping of these selective constraints and the design of experiments that allow more minimal (and quantifiable) selective constraints such that we might observe the genuine emergence of novelty, *de novo* selective causation, and life[76].

In modern physics (the prevailing conception, see Deutsch[26]), determinism is bottom-up, so the top-down constraints of the human chemist can easily be ignored. Yet, it has also been argued that one must invoke 'top-down' causation to explain causal effects in life, to be consistent with current physics[53]. This is implicated in discussions of self-reference and semantic closure as necessary to explain the life[79] and open-ended evolution[80]. However, we do not regard these features as necessary to a fundamental physics of open-ended evolution; instead, they are emergent, effective properties characterizing some of our descriptions of open-ended spaces. Concepts like 'top-down causation', 'self-reference', and 'semantic closure' introduce paradoxes associated with supervenience[54], and other paradoxes given there is no fundamental notion of 'cause' from which one might say causes are either bottom-up or top-down, or that information (semantic or otherwise) has causal power. This has placed fundamental concepts related to the nature of the living state in conflict with fundamental concepts in physics. These cannot be simply reconciled by saying life is 'emergent', as such views have made little headway in solving open scientific problems like the transition mediating the origins of life.

These paradoxes are resolved in AT. An assembly space of an object will not contain its own causation. This is important as it avoids the paradoxes of self-reference[53]. In current physics and biology, hierarchy is defined in terms of spatial scale, or energy scale (which is related to spatial extent in physics). In assembly theory, hierarchy is defined in terms of causal possibilities. The hierarchy in existence afforded by AT, and the transitions associated with this hierarchy, mean no object is contained within its own assembly space. There is no preferred spatial scale, and causation emerges from the assembly space as it expands with formation of objects[42]. The existence of objects can depend on structures that are much larger in time and space than the object itself. An example is a pencil[55]: although a pencil seems a simple, spatially small object that one can hold in one's hand, it requires a massively distributed spatiotemporal structure for its existence, e.g., in the form of a technological infrastructure spanning multiple continents across the surface of a planet. Likewise, high assembly biological molecules in high concentration (e.g. ATP concentrations found in cells) would implicate that a structure like a biological cell must also exist, along with a myriad of other non-iterable structures necessary to its existence in the high assembly index, high copy number regime of the assembly space. Our models here are simple to illustrate the core concepts, and we have shown only the first such causal transition; but we expect, given the dynamics we previously described on open ended evolution, that such causal transitions will be a recurring feature on inhabited planets that grow living and technological forms, as branching chains moving deeper into possibility and new substrates. New forms appearing along these branches will have some predictable structure as virtual objects in the assembly space are reassembled to make new objects.

**Conclusions**

In 1912, Bertrand Russell famously wrote[81] that "*The law of causality, I believe, like much that passes muster among philosophers, is a relic of a bygone age, surviving, like the monarchy,* *only because it is erroneously supposed to do no harm.*", claiming that causality was largely useless for the sciences. His view seemed consistent with the deterministic, block-universe picture of the physical sciences: if everything is just part of one grand, fixed spacetime structure, then causes and effects might be no more than a convenient fiction, an artifact of our human perspective. This vision has left some of the most fascinating questions in science underexplored and unexplained. Chief among them is the emergence of complexity, including the origin of life. Current physics describes a "block universe", with its space-time structure determined within a conception of time rooted in the relative of simultaneity of events (clocks). The physics we have outlined herein assigns a causal hierarchy to objects (and by extension events associated to these objects) introducing a concept of *causal time* defined by the conditional nature of what can exist, as defined along contingent assembly paths. Thus, in assembly theory the universe can only be described as a static block in the causal past, because events in the causal future do not yet exist (and therefore have no coordinate in spacetime geometry), see e.g. also Ellis on the idea of an evolving block universe[82,83].

The radical departure of AT from existing paradigms is the prediction of a threshold in assembly index and copy number that delineates a fundamental boundary between what structures our universe can generate spontaneously at any time, and those that can only arise due to selective construction by other objects (life). The living threshold differs depending on the causal mechanisms defining the substrate, but in combinatorial systems a threshold will always exist. Objects above the threshold, like Taxol and watches (in respective substrate-specific assembly spaces), can only exist under very specific causal circumstances that select their construction and thus the theory predicts these *cannot* form outside of an evolutionary lineage.

Assembly theory formalizes how, conditioned on the structure of what we observe to exist, there is a large space of physically possible things that could also exist, but do not, because there is no causal mechanism for their formation. Thus, in open-ended evolutionary systems, the boundary of what is possible will be expanding with the depth of causation, evolving along with the assembly of objects that exist.

The boundary of what is possible, and can come to exist, is not fixed but expands as systems become more assembled. What was once impossible becomes possible and then probable and routine once a causal mechanism is selected. No single closed algorithm defined over a predetermined alphabet or rule set can fully specify, enumerate, or predict the future set of relevant objects, even though any finite fragment of an assembly history can be simulated or analysed algorithmically after the fact. This is a physical limitation on description and prediction, unpredictability, randomness, or hypercomputation. Computation cannot be used to predict possible causal futures. This physics is testable.

Our universe is large in physical volume (approximately 93 billion lightyears in diameter) and deep in time (approximately 13.8 billion years in clock time). Less appreciated is that our universe is much larger in possibility than in realizability. The $\sim10^{80}$ atoms in the observable universe may sound vast, but these can only ever be used to construct a finite and relatively small number of objects. That is, the size of what is *possible* is far larger (an undefinable) than the size of what is *actual*. Living structures virtualize the possible within physical objects, by instantiating large amounts of causation in small volumes of space and time. In addition to our universe being deep in space and deep in time, we must contend with the fact that to support life and intelligence, we must live in a universe deep in causal possibility.

**Acknowledgements**

The development of AT has been a team effort starting since its birth in 2014. We would like to thank the following people for comments on this manuscript including Michael

Ardoline, Albert Wenger, A. J. Laucks, Michael Lachmann, Christopher Kempes, Michael Lynch, Ted Chiang, Amit Kahana, Marina Fernández Ruz, Abhishek Sharma, Keith Patarroyo. We thank the University of Glasgow, Arizona State University, NASA, the John Templeton Foundation, the Sloan Foundation, Schmidt Sciences, the Engineering and Physical Sciences Research Council, the Breakthrough Prize Foundation for funding.

**References**

1. Pasteur, M. L. New Experiments Relating to What is Termed Spontaneous Generation. *J. Cell Sci.* **2**, 118–123 (1861).
2. Carroll, S. M. Why Boltzmann Brains Are Bad. in *Current Controversies in Philosophy of Science* (Routledge, 2020).
3. Walker, S. I. *Life As No One Knows It: The Physics of Life's Emergence*. (Penguin Random House, United States of America, 2024).
4. Adami, C. What is information?†. *Philos Trans A Math Phys Eng Sci* **374**, 20150230 (2016).
5. Moynihan, T. The History of Contingency and Future-Oriented Thought. *Elements in Historical Theory and Practice* https://doi.org/10.1017/9781009358767 (2026) doi:10.1017/9781009358767.
6. Needham, J. T. A summary of some late observations upon the generation, composition, and decomposition of animal and vegetable substances; communicated in a letter to Martin Folkes Esq; President of the Royal Society, by Mr. Turbervill Needham, Fellow of the same Society. *Phil. Trans. R. Soc.* **45**, 615–666 (1748).

7. Moynihan, T. The History of Contingency and Future-Oriented Thought. *Elements in Historical Theory and Practice* https://doi.org/10.1017/9781009358767 (2026) doi:10.1017/9781009358767.
8. Reymond, J.-L., Ruddigkeit, L., Blum, L. & van Deursen, R. The enumeration of chemical space. *Wiley Interdisciplinary Reviews: Computational Molecular Science* **2**, 717–733 (2012).
9. Nicolaou, K. C., Dai, W.-M. & Guy, R. K. Chemistry and Biology of Taxol. *Angewandte Chemie International Edition* **33**, 15–44 (1994).
10. Kanda, Y. *et al.* Two-Phase Synthesis of Taxol. *J. Am. Chem. Soc.* **142**, 10526–10533 (2020).
11. Nicolaou, K. C., Riemer, C., Kerr, M. A., Rideout, D. R. & Wrasidlo, W. W. Design, synthesis and biological activity of protaxols. *Nature* **364**, 464–466 (1993).
12. Paley, W. *Natural Theology: Or, Evidences of the Existence and Attributes of the Deity Collected from the Appearances of Nature*. (R. Faulder, London, 1802).
13. Cortês, M., Kauffman, S. A., Liddle, A. R. & Smolin, L. The TAP equation: Evaluating combinatorial innovation. *European Economic Review* **179**, 105144 (2025).
14. Simon, H. A. The Architecture of Complexity. *Proceedings of the American Philosophical Society* **106**, 467–482 (1962).
15. Krakauer, D. C. *The Complex World: An Introduction to the Foundations of Complexity Science*. (SFI Press, 2024).

16. Darwin, C. *The Origin of Species by Means of Natural Selection, or the Preservation of Favoured Races in the Struggle for Life*. (John Murray, London, 1859).

17. Wagner, A. *Arrival of the Fittest: Solving Evolution's Greatest Puzzle*. (Penguin Random House, 2014).

18. Smith, J. M. & Szathmary, E. *The Major Transitions in Evolution*. (OUP Oxford, 1997).

19. Smith, E. & Morowitz, H. *The Origin and Nature of Life on Earth: The Emergence of the Fourth Geosphere*. (Cambridge University Press, 2016).

20. Walker, S. I. AI Is Life. *Noema Magazine* (2023).

21. Arthur, W. B. *The Nature of Technology: What It Is and How It Evolves*. (Simon and Schuster, 2009).

22. Sharma, A. *et al.* Assembly theory explains and quantifies selection and evolution. *Nature* **622**, 321–328 (2023).

23. Marshall, S. M. *et al.* Identifying molecules as biosignatures with assembly theory and mass spectrometry. *Nat Commun* **12**, 3033 (2021).

24. Marletto, C. Constructor theory of life. *Journal of The Royal Society Interface* **12**, 20141226 (2015).

25. *Theory of Self-Reproducing Automata*. (University of Illinois, Urbana IL, 1966). doi:10.21236/AD0688840.

26. Deutsch, D. Constructor theory. *Synthese* **190**, 4331–4359 (2013).

27. Shang, H. A generic hierarchical model of organic matter degradation and preservation in aquatic systems. *Commun Earth Environ* **4**, 16 (2023).

28. Bowman, J. C., Petrov, A. S., Frenkel-Pinter, M., Penev, P. I. & Williams, L. D. Root of the Tree: The Significance, Evolution, and Origins of the Ribosome. *Chem. Rev.* **120**, 4848–4878 (2020).

29. Marshall, S. M., Moore, D. G., Murray, A. R. G., Walker, S. I. & Cronin, L. Formalising the Pathways to Life Using Assembly Spaces. *Entropy* **24**, 884 (2022).

30. Jirasek, M. *et al.* Investigating and Quantifying Molecular Complexity Using Assembly Theory and Spectroscopy. *ACS Cent. Sci.* **10**, 1054–1064 (2024).

31. Patarroyo, K. Y. *et al.* Quantifying the Complexity of Materials with Assembly Theory. Preprint at https://doi.org/10.48550/arXiv.2502.09750 (2025).

32. Patarroyo, K. Y., Sharma, A., Walker, S. & Cronin, L. AssemblyCA: A Benchmark of Open-Endedness for Discrete Cellular Automata. in *Second Agent Learning in Open-Endedness Workshop* (2023).

33. Janin, E., Shkolnik, E., Slocombe, L. & Walker, S. Searching for Life as We Don't Know It: Detecting Signatures of Chemical Selection in Exoplanet Atmospheres. in *246th Meeting of the American Astronomical Society* vol. 246 313.06 (2025).

34. Seet, I., Patarroyo, K. Y., Siebert, G., Walker, S. I. & Cronin, L. Rapid Computation of the Assembly Index of Molecular Graphs. Preprint at https://doi.org/10.48550/arXiv.2410.09100 (2024).

35. Walker, S. I. & Davies, P. C. W. The 'Hard Problem' of Life. in *From matter to life: information and causality* 19–37 (Cambridge University Press, 2017).

36. Gisin, N. Indeterminism in Physics, Classical Chaos and Bohmian Mechanics: Are Real Numbers Really Real? *Erkenn* **86**, 1469–1481 (2021).

37. Gisin, N. Mathematical languages shape our understanding of time in physics. *Nat. Phys.* **16**, 114–116 (2020).

38. Wigner, E. P. Invariance in Physical Theory. *Proceedings of the American Philosophical Society* **93**, 521–526 (1949).

39. Liu, Y. *et al.* Exploring and mapping chemical space with molecular assembly trees. *Science Advances* **7**, eabj2465 (2021).

40. Marletto, C. *et al.* Emergence of Constructor-Based Irreversibility in Quantum Systems: Theory and Experiment. *Phys. Rev. Lett.* **128**, 080401 (2022).

41. Noether, E. Invariant variation problems. *Transport Theory and Statistical Physics* **1**, 186–207 (1971).

42. Walker, S. I., Mathis, C., Marshall, S. & Cronin, L. Experimental Measurement of Assembly Indices are Required to Determine The Threshold for Life. Preprint at https://doi.org/10.48550/arXiv.2406.06826 (2024).

43. Anderson, P. W. More Is Different. *Science* **177**, 393–396 (1972).

44. Hawking, S. *The Illustrated A Brief History of Time: Updated and Expanded Edition*. (Random House Publishing Group, 1996).

45. Jacobs, J. D. *Causal Powers*. (Oxford University Press, 2017).

46. Lloyd, S. Measures of Complexity a non--exhaustive list. *IEEE Control Systems Magazine* **21**, 7–8 (2001).

47. Reichenbach, H. *The Theory of Probability*. (University of California Press, 1971).

48. Venn, J. *The Logic of Chance: An Essay on the Foundations and Province of the Theory of Probability, with Especial Reference to Its Logical Bearings and Its Application to Moral and Social Science, and to Statistics*. (Macmillan, 1888).

49. Mises, R. V. *Probability, Statistics, and Truth*. (Courier Corporation, 1981).

50. Popper, K. R. The Propensity Interpretation of Probability. *The British Journal for the Philosophy of Science* **10**, 25–42 (1959).

51. Bayes, T. An essay towards solving a problem in the doctrine of chances. *MD Comput* (1991).

52. Jeffreys, H. *The Theory of Probability*. (OUP Oxford, 1998).

53. Williamson, T. *Modal Logic as Metaphysics*. (OUP Oxford, 2013).

54. Lewis, D. *On the Plurality of Worlds*. (Blackwell, 1986).

55. Kripke, S. A. *Naming and Necessity*. (Blackwell, 1980).

56. Bird, A. *Nature's Metaphysics: Laws and Properties*. (OUP Oxford, 2007).

57. Mumford, S. & Anjum, R. L. *Getting Causes from Powers*. (OUP Oxford, 2011).

58. DeLanda, M. *Intensive Science and Virtual Philosophy*. (Bloomsbury Publishing, 2013).

59. Deleuze, G. *Difference and Repetition*. (Cambridge University Press, 1994).

60. Pearl, J. *Causality*. (Cambridge University Press, 2009).

61. Marletto, C. *The Science of Can and Can't: A Physicist's Journey through the Land of Counterfactuals*. (Viking Press, 2021).

62. Cronin, L., Pagel, S. & Sharma, A. Chemputer and Chemputation -- A Universal Chemical Compound Synthesis Machine. Preprint at https://doi.org/10.48550/arXiv.2408.09171 (2025).

63. Mastrogiovanni, A. M. The Descent of Abstraction: Iterability, Assembly Theory, and the Origin of Ideality. *Philosophy Today* **69**, 427–451 (2025).

64. Walker, S. I. & Davies, P. C. W. The algorithmic origins of life. *Journal of The Royal Society Interface* **10**, 20120869 (2013).

65. Falcon, A. Aristotle on Causality. in *The Stanford Encyclopedia of Philosophy* (eds Zalta, E. N. & Nodelman, U.) (Metaphysics Research Lab, Stanford University, 2023).

66. Ardoline, M. J. Life Against Smallism: Assembly Theory, Scale, and the Order of Explanation. *Philosophy Today* **69**, 453–467 (2025).

67. Jaynes, E. T. *Probability Theory: The Logic of Science*. (Cambridge University Press, 2003).

68. Fox, G. Origin and Evolution of the Ribosome. *Cold Spring Harbor Perspectives in Biology* **2**, a003483 (2010).

69. Thomson, W. On an absolute thermometric scale founded on Carnot's theory of the motive power of heat and calculated from Regnaut's observations. *Cambridge Philosophical Society Proceedings* **1**, 100–106 (1848).

70. Lössl, P., van de Waterbeemd, M. & Heck, A. J. The diverse and expanding role of mass spectrometry in structural and molecular biology | The EMBO Journal. *EMBO Journal* **35**, 2634–2657 (2016).

71. Jarzynski, C. Nonequilibrium Equality for Free Energy Differences. *Phys. Rev. Lett.* **78**, 2690–2693 (1997).

72. Crooks, G. E. Path-ensemble averages in systems driven far from equilibrium. *Phys. Rev. E* **61**, 2361–2366 (2000).

73. Kauffman, S. A. *The Origins of Order: Self-Organization and Selection in Evolution*. (Oxford University Press, 1993).

74. Hirakawa, Y. *et al.* Interstep compatibility of a model for the prebiotic synthesis of RNA consistent with Hadean natural history. *Proceedings of the National Academy of Sciences* **122**, e2516418122 (2025).

75. Richert, C. Prebiotic chemistry and human intervention. *Nat Commun* **9**, 5177 (2018).

76. Cooper, G. J. T., Walker, S. I. & Cronin, L. A Universal Chemical Constructor to Explore the Nature and Origin of Life. in *Conflicting Models for the Origin of Life* 101–130 (John Wiley & Sons, Ltd, 2023). doi:10.1002/9781119555568.ch6.

77. Walton, C., Rimmer, P. B., Williams, H. & Shorttle, O. Prebiotic Chemistry in the Wild: How Geology Interferes with the Origins of Life. Preprint at https://doi.org/10.26434/chemrxiv.13198205.v1 (2020).

78. Cronin, L. & Walker, S. I. Beyond prebiotic chemistry. *Science* **352**, 1174–1175 (2016).

79. Pattee, H. H. Evolving Self-reference: Matter, Symbols, and Semantic Closure. in *Laws, Language and Life: Howard Pattee's classic papers on the physics of symbols with contemporary commentary* (eds Pattee, H. H. & Rączaszek-Leonardi, J.) 211–226 (Springer Netherlands, Dordrecht, 2012). doi:10.1007/978-94-007-5161-3_14.

80. Sayama, H. Construction theory, self-replication, and the halting problem. *Complexity* **13**, 16–22 (2008).

81. Russell, B. On the Notion of Cause. *Proceedings of the Aristotelian Society* **13**, 1–26 (1912).

82. Ellis, G. F. R. The evolving block universe and the meshing together of times. *Annals of the New York Academy of Sciences* **1326**, 26–41 (2014).

83. Ellis, G. Time really exists! The evolving block universe. **7**, (2014).

**Definitions**

**Assemblage:** The accumulated causal possibilities, recursively assembled to generate the set of distinguishable objects and their copies. This is a causally deep representation of matter, as it constitutes all the virtual structures necessary to the existence of the observed objects.

**Assembly, A:** The density of causal possibilities per observed object, taken as the ratio of the copy number in the virtual space of objects to total number of countable objects in a sample. This captures cumulative selection in accounting for both the causal depth of objects in a combinatorially explosive space (assembly index) and the persistence of the mechanisms to produce these objects (copy number)

**Assembly index, $a_i$:** The minimal number of causal joins required to realise an object under a specified set of generative rules determined from the object's intrinsic causes. This is intensive to the object.

**Assembly path:** A finite sequence of causal joins terminating on an observed object.

**Assembly space:** The physical space denoting all causal possibilities, and subject to metrological interrogation.

**Assembly time, $d$:** The number of causal joins in a constructive process to produce an object.

**Cause:** Structure that appears on a minimal assembly path to an observed object. These structures are causal joins of two fragments, they are not *a priori* labelled, and their precise identity need not be known.

**Causation:** A material attribute revealed in the existence of objects, as encoded in a minimal assembly path.

**Contingency:** The dependence of an object's existence on a hierarchy of objects causally prior to its existence, where each object carries this dependence in a virtual space of finite support (a minimal assembly path).

**Copy number, $n_i$:** The countable number of each distinguishable object type, $i$, (defined up to limits of measurement), which provides evidence of a contingent causal chain necessary to its existence.

**Epistemological Threshold, $a_M$:** A defined boundary in assembly space, set by the resolution of the measurement limit, $M$, where objects with a larger assembly index cannot exist and be observed in the absence of selection. This defines the boundary of the abiotic assembly space and is useful for detection of living and intelligent lineages.

**Measurement limit, $M$:** The minimum copy number of a distinguishable object necessary for that object to be reliably detected and identified by a given measurement or observational apparatus.

**Possibility:** A causal join that can happen, or an object that can exist. These are encoded in the assembly space: possible objects or events are those that can be generated by selection.

**Probability:** A phenomenological means to encode the role of the environment in an object's existence, which allows quantifying the production rate or error-rate of making a specific object with a given accuracy in each existential context.

**Object, $o_i$:** A physical entity that is (1) countable, (2) finite, and (3) can be disassembled (or assembled) by a finite sequence of recursive steps. Measurement places fundamental limits on the observation of objects, and their decomposition into parts.

**Ontological Threshold, $a_1$:** A defined boundary in assembly space, set where the resolution of measurement is *exactly one object*, and the threshold $a_M$ becomes an *ontological limit* delineating the maximum assembly index at which an object can exist at least once in a finite physical system in the absence of selection.

**Selection:** The action of causal mechanisms allowing only subsets of possible structures to persist and be observed. Unlike biological selection, which requires heredity and variation, this more general definition of selection includes biological mechanisms and those that happen outside of biology, such as the causal selective processes that led to the emergence of biological life.

**Virtual object:** Structures with causal power that appear on a minimal assembly path and these may not be independently observable, because there does not necessarily exist a causal mechanism that can produce identical, countable copies of these structures in high numbers.

## Appendix A: Derivation of assembly index threshold, $a_M$

An assembly theoretic question of interest is: *What is the limit on the space of possible objects with no selection?* The assembly space is a recursive, branching space, where fragments are combined to produce new structure, leading to an increasing number of branches with assembly steps. As a simplification, we consider a branching process where each node has $b$ descendants. This branching structure is a feature of the assembly space, and $b$ allows encoding what is possible in terms of causal steps from each fragment in the space. In reality, $b$ is not a global parameter, but instead reflects local structure in the assembly space. For example, $b$ might be higher in regions of chemical space where there is a dominance of C atoms, and lower where there is a dominance of N atoms because of the possible bonds available to support causal joining in these regions.

For purposes of illustrating the existence of a threshold, we treat $b$ as global (we will treat locality in the topology of the assembly space in future work). We consider the case of equivalent exploration of everything that could be possible, that is, no selection: all branches at a given depth are equivalent and possible. In a combinatorial, recursively constructable space, the number of possible objects formed in successions of such steps is expected to grow exponentially: what we present here is a very approximate argument as the space will grow much faster. The simplifications do not impact the conclusion of an objectively determinable boundary for the assembly index, $a_M$, but merely serves to elucidate its existence.

In the absence of selection, all branches are realized. We assume a branching space with resource splitting among the branches such that the copy number at a node $i$ in the branching structure, after $d$ constructive steps, has the form:

**(A.1)**

$$n_i(d) = \frac{N_T}{(1+b)^{d+1}}$$

Here, $d$ is the depth counting the number of assembly steps, $b$ is the branching, and $N_T$ is the total number of objects. The form of Eq. A.1 captures how the copy number per node decreases with steps at a rate driven by the branching factor. We may equivalently write this as

$$n_i(d) = N_T \, e^{-(d+1)\ln(1+b)}$$

Which highlights the logarithmic dependency on the branching factor in the exponential drop-off of copy number with increasing steps.

We define the living threshold at causal depth $a_M$, where the condition at the limit of observation (epistemological) or existence (ontological) is satisfied, e.g., where

**(A.2)**

$$n(a_M) < M$$

Here $M$ is the epistemological bound arising due to measuring instrument resolution, and when M = 1 we recover the ontological bound. Solving for the threshold using equations B.1 and B.2:

$$\frac{N_T}{(1+b)^{a_M+1}} = M$$

$$(1+b)^{a_M+1} = \frac{N_T}{M}$$

$$(a_M+1)\ln(1+b) = \ln\left(\frac{N_T}{M}\right)$$

which yields a maximum causal depth for observed objects with assembly index:

**(A.3)**

$$a_M = \left\lfloor \frac{\ln\left(\frac{N_T}{M}\right)}{\ln(1+b)} \right\rfloor - 1$$

Since $a_M$ represents a conservative upper bound on the number of steps to observe an object, it follows that objects with assembly indices $a_i > a_M$ should not be observable in the absence of a mechanism of selection culling branching possibilities (since $a_i$ represents the minimum number of causal joins for the existence of the object).

Equation **B.3** corresponds to the copy number at each branching node, the total copy number in the layer of objects formed after $d$ causal steps, $N(d)$, can be found by multiplying by the number of nodes after d steps, e.g. $N(d) = \sum n_i(d) = b^d\, n_i(d)$, yielding

**(A.4)**

$$N(d) = b^d\, N_T\, e^{-(d+1)\ln(1+b)}$$

A few notes about the approximations we used to simplify the math. The two conditions of (1) combinatorial explosion with increasing combinatorics and (2) finite resource are sufficient to guarantee a bound in the absence of selection. The exact functional form of $n_i(d)$ is not critical to our argument; in general, we expect $n_i(d) \propto 1/b^d$, such that copy number decreases exponentially with steps due to explosive growth in the number of combinatorial possibilities with each successive step. Herein, our estimates are conservative as we expect branching to increase with assembly index.

As a minor technical note we did not impose a sharp cutoff for $n_i(d)$ at $d = a_M$ for the approximations presented here, and some (unphysical) residual mass is still distributed in the region $d > a_M$ (e.g., the number density the nodes at $d = a_M$ would distribute to their descendants but this in unphysical as it is not enough to make a single copy of these descendants). A more physical (rescaled) treatment could redistribute the residual mass upward to branches with $d < a_M$, in practice this would lead to a small correction since generally for any physical system of interest the residual mass would be $<< N_T$.

**Appendix B: Derivation of abiotic assembly**

We can use equation **A.3** and **A.4** to solve for abiotic assembly, $A_{abiotic}$, starting from the equation for assembly, $A$:

$$A = \frac{1}{N_T} \sum_{i \epsilon n_i > 1}^{N} n_i \, e^{a_i}$$

Summing over all objects with the same assembly index (using Eq. **A.4**) yields:

**B.1**

$$A_{abiotic} = \sum_{1}^{a_M} e^{a} \, \frac{\mathrm{b}^{a} \, N_T}{(1+b)^{a+1} N_T}$$

Or equivalently

**B.2**

$$A_{abiotic} = \frac{1}{(1+b)} \sum_{1}^{a_M} \left(\frac{eb}{1+b}\right)^{a}$$

This is a finite geometric series. Using $\sum_{1}^{a_M} \mathrm{r}^{a} = \frac{r\,(1-r^{a_M})}{1-r}$, with $\mathrm{r} = \frac{eb}{1+b}$ yields:

$$\sum_{1}^{a_M} \left(\frac{eb}{1+b}\right)^{a} = \frac{eb\left(1-\left(\frac{eb}{1+b}\right)^{a_M}\right)}{(1+b)\left(1-\left(\frac{eb}{1+b}\right)\right)}$$

$$= \frac{eb\left(1-\left(\frac{eb}{1+b}\right)^{a_M}\right)}{(1+b-eb)}$$

Using Eq. B.3 with $a_M = \left\lfloor \frac{\ln\left(\frac{N_T}{M}\right)}{\ln(1+b)} \right\rfloor - 1$ , we need to evaluate:

$$\left(\frac{eb}{1+b}\right)^{a_M} = \left(\frac{eb}{1+b}\right)^{\left\lfloor \frac{\ln\left(\frac{N_T}{M}\right)}{\ln(1+b)} \right\rfloor - 1}$$

$$= \left(\frac{1+b}{eb}\right)\left(\frac{eb}{1+b}\right)^{\left\lfloor \frac{\ln\left(\frac{N_T}{M}\right)}{\ln(1+b)} \right\rfloor}$$

Using $x^{\ln y} = e^{\ln y \ln x} = y^{\ln x}$ yields:

$$= \left(\frac{1+b}{eb}\right)\left(\frac{N_T}{M}\right)^{\ln\left(\frac{eb}{1+b}\right)/\ln(1+b)}$$

Substituting for the sum in **C.2** yields:

$$A_{abiotic} = \frac{1}{(1+b)}\left(\frac{1+b}{eb}\right)\left(\frac{N_T}{M}\right)^{\ln\left(\frac{eb}{1+b}\right)/\ln(1+b)}$$

$$A_{abiotic} = \frac{1}{eb}\left(\frac{N_T}{M}\right)^{\ln\left(\frac{eb}{1+b}\right)/\ln(1+b)}$$

In the main text we assume $N_T \gg 1$ and focus on the scaling exponent, which can be simplified to:

$$\ln\left(\frac{eb}{1+b}\right)/\ln(1+b) = \frac{1+\ln b}{\ln(1+b)} - 1$$

**B.3**

$$A_{abiotic} \propto N_T^{\frac{1+\ln b}{\ln(1+b)}-1}$$

The exponent $\frac{1+\ln b}{\ln(1+b)} - 1 < 1$ for all $b > 0$ so $A$ grows as a sub-linear power law in $N_T$ and $A$ is extensive.